\documentclass[aps,amsmath,twocolumn,prd,superscriptaddress,nofootinbib]{revtex4}
\pdfoutput=1
\usepackage{graphicx}
\usepackage{amsfonts}
\usepackage{amssymb}
\usepackage{amsbsy}
\usepackage{amsmath}
\usepackage{latexsym}
\usepackage{natbib}
\usepackage{bm}
\usepackage{subfigure}
\usepackage{color}
\usepackage{hyperref}
\usepackage{lscape}
\usepackage{ifthen}
\usepackage{xstring}

\DeclareFontFamily{OT1}{rsfs}{} 
\DeclareFontShape{OT1}{rsfs}{m}{n}{<-7> rsfs5 
    <7-10> rsfs7 <10-> rsfs10}{}   
\DeclareMathAlphabet{\scr}{OT1}{rsfs}{m}{n}

\newcommand*{\di}{\partial}

\def\be {\begin{equation}}
\def\ee  {\end{equation}}
\def\bea {\begin{eqnarray}}
\def\eea {\end{eqnarray}}
\def\nn {\nonumber}

\renewcommand*{\k}{\hat{k}}

\newcommand*{\Pl}{\text{Pl}}

\newcommand*{\nmax}{n_{\text{max}}}

\newcommand*{\M}{M_{\star}}

\def\k{\mathbf{k}}
\def\x{\mathbf{x}}
\def\y{\mathbf{y}}

\begin{document}

\title{Primordial polymer perturbations}

\author{Sanjeev S.\ Seahra}
 
\affiliation{Department of Mathematics and Statistics, University of New Brunswick, Fredericton, NB, Canada E3B 5A3} 

\author{Iain A.\ Brown}

\affiliation{Institute of Theoretical Astrophysics, University of Oslo, P.O. Box 1029 Blindern, N-0315 Oslo, Norway}

\author{Golam Mortuza Hossain}

\affiliation{Department of Physical Sciences, Indian Institute of Science Education and Research Kolkata, Mohanpur Campus, PO: Krishi Viswavidyalaya, Nadia - 741 252, WB, India}

\author{Viqar Husain}

\affiliation{Department of Mathematics and Statistics, University of New Brunswick, Fredericton, NB, Canada E3B 5A3} 
 

\begin{abstract}

We study the generation of primordial fluctuations in pure de Sitter inflation where the quantum scalar field dynamics are governed by polymer (not Schr\"odinger) quantization.  This quantization scheme is related to, but distinct from, the structures employed in Loop Quantum Gravity; and it modifies standard results above a polymer energy scale $M_{\star}$.  We recover the scale invariant Harrison Zel'dovich spectrum for modes that have wavelengths bigger than $M_{\star}^{-1}$ at the start of inflation.  The primordial spectrum for modes with initial wavelengths smaller than $M_\star^{-1}$ exhibits oscillations superimposed on the standard result.  The amplitude of these oscillations is proportional to the ratio of the inflationary Hubble parameter $H$ to the polymer energy scale.  For reasonable choices of $\M$, we find that polymer effects are likely unobservable in CMB angular power spectra due to cosmic variance uncertainty, but future probes of baryon acoustic oscillations may be able to directly constrain the ratio $H/\M$.

\end{abstract}

\maketitle

\section{Introduction}

It is commonly believed that quantum gravity effects may significantly alter ``standard'' physics near the Planck scale.  For example, string theory posits that extra dimensions with compact topology will become visible at energies approaching $M_\Pl$, while loop quantum gravity asserts that continuous classical spacetime is replaced by quantum spin networks on small scales.  Unfortunately, the huge discrepancy between the Planck scale and typical energies in the nearby universe make it virtually impossible to experimentally or observationally test such ideas.  In fact, the only available data comes indirectly through measurements of the cosmic microwave background: If one accepts the inflationary paradigm, this thermal relic of the hot big bang depends on the spectrum of primordial perturbations generated when the temperature of the universe was a few orders of magnitude less than the Planck scale.  This makes inflation the only known phenomenon which both involves Planckian energies and has measurable 
consequences for the observable universe.  It is therefore a crucial issue for theories of quantum gravity to predict how ``new physics'' near the Planck scale affects the spectrum of primordial perturbations generated from inflation.

A useful way of thinking about the effects of new physics in inflation involves viewing the quantum generation of inflationary fluctuations as a trans-Planckian problem \cite{Martin:2000xs}.   The idea is as follows:  if one takes the very largest scale cosmological perturbations which are relevant for observations and tracks them backwards in time, one finds that at some finite time during the inflationary epoch their physical wavelengths will become smaller than the Planck length.  Hence, the early time evolution of such modes will necessarily be sensitive to any new physics manifest at small scales, which then implies that there should be some imprint of very high energy phenomena on very large cosmological distances.  Of course, the key questions are the amplitude and nature of these effects, which in turn depend on the nature of the small scale modification.  

In the literature, various authors have considered \emph{ad hoc} modifications to scalar field dispersion relations \cite{Brandenberger:2000wr,Martin:2000xs,Niemeyer:2000eh,Shankaranarayanan:2002ax}, due to non-commutativity \cite{Chu:2000ww,Lizzi:2002ib,Brandenberger:2002nq,Hassan:2002qk}, or modified uncertainty relations \cite{Kempf:2000ac,Easther:2001fi,Kempf:2001fa,Ashoorioon:2004vm,Kempf:2006wp}.  There have been attempts also to calculate trans-Planckian contributions to the primordial power spectrum in a model-independent way by imposing initial conditions on a ``new physics hyeprsurface'' \cite{Niemeyer:2002kh,Bozza:2003pr,Easther:2002xe,Danielsson:2002kx}.  Recently, effects arising in Horava-Lifshitz gravity have been reported \cite{Ferreira:2012xa}.  A feature of many (but not all) of these studies is that short distance effects superimpose oscillations on the conventional scale-invariant power spectrum with amplitude $(H/\M)^{\gamma}$, where $\M$ is the energy threshold above which the modifications are 
important, $H$ is the inflationary Hubble parameter, and the power $\gamma \gtrsim 1$ depends on the model.

In this paper, we explore a different class of ``new physics'' suggested by the ``background independent'' (or ``polymer'') approach to quantization that is deployed in loop quantum gravity (LQG) \cite{Thiemann:2007zz}.  In this programme, classical geometric variables such as the metric are represented at the quantum level by graphs on spatial 3-manifolds known as spin networks. Fundamental geometric information such as areas, volumes and their evolution are encoded by densitized triads and holonomies of connection 1-forms over the edges of these graphs. The key point is that while quantum operators corresponding to the holonomies are well-defined, operators corresponding to the connection 1-forms themselves are not. This implies the Hilbert space of LQG has distinct properties from the standard one underlying Schr\"odinger quantum mechanics and quantum field theory.

The novel features of this approach to quantization are best illustrated by considering ordinary quantum mechanics \cite{Ashtekar:2002sn}:  Consider a particle moving in one dimension and described by a position $x$ and its conjugate momentum $p$.  In the conventional Schr\"odinger quantization (SQ) of the system the particle's state is described by an element of a Hilbert space in which the action of the position $\hat{x}$ and momentum operators $\hat{p}$ are well-defined.  On the other hand, an LQG-inspired quantization makes use of an alternative Hilbert space where the operator corresponding to $p$ is not defined, but the associated ``holonomy'' operator is.  Since $p$ is the generator of infinitesimal translations, the appropriate identification of its holonomy is the \emph{finite} translation operator $\hat{U}_{\lambda}$ (whose action is to displace the particle by a distance $\Delta x = -\lambda$).  The quantization algorithm based on this Hilbert space is called ``polymer quantization'' (PQ) since it is motivated 
by the spin network structure of LQG, where the excitations of the gravitational field occur along the edges of a graph; i.e. they are one-dimensional like a polymer.\footnote{It is important to note that PQ has been proposed as an alternative to standard quantum theory independently of any quantum gravity considerations:  In particular, Halvorson \cite{Halvorson} proposed such a quantization as an alternative to SQ that allowed for normalizable position eigenstates in the Hilbert space.  The existence of such states implies that the standard uncertainty principle does not hold in the polymer picture \cite{Hossain:2010wy}.}

The lack of a natural momentum operator in PQ may seem alarming, but one can easily define an effective $\hat{p}$ by constructing a simple finite difference stencil for the Schr\"odinger momentum operator $i\di_{x}$ using finite translations $\hat{U}_{\lambda_{\star}}$.  The characteristic size of this stencil $\lambda_{\star}$ is  an arbitrary fixed parameter of the quantization that defines a polymer energy scale $\M$.  We expect to recover ordinary SQ at energies less than $\M$ (since our finite difference approximation to $\hat{p}$ will be very good in that regime) while at higher energies we would expect the predictions of PQ and SQ to differ substantially.

This expectation has been explicitly confirmed by calculating the spectrum of a polymer quantized simple harmonic oscillator of mass $m$ \cite{Ashtekar:2002sn,Corichi:2007tf,Hossain:2010eb}.  One finds that the energies $E_{n}$ of eigenstates of the Hamiltonian approximate the well-known SQ values when $m E_{n} / \M^{2} \ll 1$.  Similar results are available for the Coulomb \cite{Husain:2007bj}, inverse-square \cite{Kunstatter:2008qx}, and other \cite{Kunstatter:2009ua,Kunstatter:2010fa} spherical potentials; though the issue of boundary conditions at the origin of spherical coordinates must be handled carefully \cite{Kunstatter:2012ra}.  The limit in which one can obtain reasonable polymer approximations to Schr\"odinger wavefunctions has also been considered \cite{Fredenhagen:2006wp}.

Polymer effects have also been studied extensively in the context of quantum cosmology \cite{Bojowald:2001xe,Ashtekar:2003hd,lrr-2005-11}.  Specifically, the polymer treatment of geometric quantities in FRW models preserves the predictions of general relativity at low curvature while replacing the big bang singularity with a big bounce when the density of the universe is $\sim \M^{4}$ \cite{Ashtekar:2006rx}.  Conversely, polymer quantization of a homogeneous and massless scalar in the early universe has been shown to replace the big bang with a past eternal de Sitter phase with Hubble parameter $H \sim \M^{2}/M_{\Pl}$ \cite{Hossain:2009ru}.
 
It is natural to try to extend these results from polymer quantum mechanics to quantum field theory.  In that vein, the PQ of a scalar field in Minkowski space has been considered \cite{Ashtekar:2002vh}, assuming compact topology \cite{Laddha:2010hp}, using semi-classical approximations \cite{Hossain:2009vd}, via an effective spatial lattice \cite{Husain:2010gb}, and by direct quantization of Fourier modes \cite{Hossain:2010eb}.  In the last approach, it was shown that Fourier modes with $e^{i\k\cdot \x}$ spatial dependence exhibit exotic polymer behaviour if $|\k| \gg \M$.  

In this paper, we use the techniques introduced in \cite{Hossain:2010eb} to study the PQ of a scalar field in a de Sitter inflationary universe.  The motivation is obvious: since the physical wavenumber of a given Fourier mode is inversely proportional to the scale factor in an expanding universe, its behaviour will be dominated by polymer effects in the asymptotic past.  Hence, we would expect that the PQ of a scalar field during inflation will result in potentially observable modifications to the primordial perturbation spectrum.  The current work confirms and quantifies this expectation.

The organization of the paper is as follows:  In \S\ref{sec:pert review} we recall the standard textbook calculation of the primordial power spectrum as well as an alternative formulation based on quantization of individual Fourier modes.  In \S\ref{sec:SQ} we describe how mode-by-mode quantization is achieved in the standard Schr\"odinger picture, while in \S\ref{sec:PQ} we present the calculation in the polymer formalism.  In \S\ref{sec:power spectrum} we present numerical and semi-analytic results for the polymer primordial spectrum, and in \S\ref{sec:observations} compare them to observations of the cosmic microwave background and large scale structure of the universe.  We summarize and discuss our main results in \S\ref{sec:discussion}.  The appendices give a technical introduction to polymer quantum mechanics \S\ref{sec:polymer}, list cosmological scaling relations used throughout the paper \S\ref{sec:scaling}, and derive some technical formulae \S\ref{sec:k star derivation}.

\section{Generation of primordial perturbations in a de Sitter universe}\label{sec:pert review}

In this section, we review the calculation of the spectrum of primordial perturbations in a de Sitter inflationary universe using two complementary methods:  The first is based on the quantization of the scalar field in real space and the subsequent Fourier decomposition of the quantum operators.  The generation of fluctuations follows from the fact that quantum operators obey the classical equations of motion; i.e.; this approach uses the Heisenberg picture.  The second method involves first Fourier decomposing the field, which reduces the system to a collection of independent oscillators with time dependent parameters, and then quantizing each oscillator.  In this case, the generation of fluctuations follows from the solution of the resulting one-dimension Schr\"odinger equation with a time dependent mass and potential for the wavefunction (i.e., this approach uses the Schr\"odinger picture).  We will make use of the latter approach when considering the polymer quantization of the scalar field.

\subsection{Quantization in real space}

We consider a massless scalar field propagating in a de Sitter background:\footnote{Recall that in a de Sitter background, the equation of state $\rho + p = 0$ implies that metric perturbations are decoupled from scalar field fluctuations, so $\phi$ is automatically a gauge-invariant quantity.}
\begin{equation}
ds^{2} = -dt^{2}+a^{2}(t) d\x^{2}, \quad a(t) = \exp Ht.
\end{equation} 
The Hamiltonian of the scalar field is given by
\begin{equation}\label{eq:standard H}
{\scr H}_{\phi} = \int d^{3} x \, a^{3}\left[ \frac{1}{2a^{6}} \pi^{2}+ \frac{1}{2a^{2}}  (\mathbf{\nabla}\phi)^{2} \right].
\end{equation}
Here, $\pi$ is the momentum conjugate to $\phi$ such that $\{\phi(t,\x),\pi(t,\y)\} = \delta^{(3)}(\x-\y)$.  To quantize this system, one customarily promotes both $\phi$ and $\pi$ to operators and imposes the commutation relations $[\hat\phi(t,\x),\hat{\pi}(t,\y)] = i\delta^{(3)}(\x-\y) $.  Then, the field operator is decomposed into Fourier modes via
\begin{multline}
	\hat{\phi}(t,\x) = \frac{1}{\sqrt{V_{0}}}\sum_{\k}\hat{\phi}_{\k}(t) e^{i {\k}\cdot{\x}}  \\ =  \frac{1}{\sqrt{V_{0}}} \sum_{\k} \left[ f_{\k}(t) e^{-i\k \cdot \x}\hat{a}_{\k} + f^{*}_{\k}(t) e^{+i\k \cdot \x} \hat{a}^{\dag}_{\k} \right],
\end{multline}
where $V_{0}$ is the fiducial volume (with respect to the flat 3-metric $diag(1,1,1)$) used in our box normalization i.e.
\begin{equation}
V_{0} = \int d^{3}x.
\end{equation}
In the Heisenberg picture, $\hat\phi$ satisfies the classical equation of motion $\Box\hat\phi = 0$, which implies that
\begin{equation}\label{eq:classical EOM}
	\ddot{f}_{\k} + 3H \dot{f}_{\k} + \frac{k^{2}}{a^{2}} f_{\k} = 0.
\end{equation}
The ladder operators satisfy the commutation relation $[ \hat{a}_{\k},\hat{a}_{\k}^{\dag} ] = 1$.  During inflation, we assume that the field is in the vacuum state annihilated by the $\hat{a}_{\k}$ operators; i.e.\ $\hat{a}_{\k}|\psi\rangle = 0$.  Then, the power spectrum of primordial perturbations generated during inflation is 
\begin{equation}
	\mathcal{P}_{\phi}(k) = \frac{k^{3}}{2\pi^{2}}  \langle \phi_{\k}^{2} \rangle\bigg|_{k \ll aH}, \quad \langle \phi_{\k}^{2} \rangle = \langle \psi | \hat{\phi}_{\k}^{2} | \psi \rangle = |f_{\k}|^{2}.
\end{equation}
Hence, we see that the power spectrum is entirely determined by solution of the classical equation of motion (\ref{eq:classical EOM}).  Now, there are infinitely many solutions of the ODE and each particular solution will pick out a different quantum vacuum state; i.e., there is an ambiguity in the calculation.  This is commonly resolved by demanding that the Minkowski vacuum state is recovered in the appropriate limit of the parameter space, which picks out the solution
\begin{equation}
	f_{\k} = \frac{H}{\sqrt{2k^{3}}} \left(1 - i\frac{k}{Ha} \right) e^{ik/Ha}, \quad \mathcal{P}_{\phi}(k) = \left( \frac{H}{2\pi} \right)^{2}.
\end{equation}
This choice is referred to as the Bunch-Davies or adiabatic vacuum, and it gives rise to the familiar scale invariant Harrison-Zel'dovich (HZ) spectrum $\mathcal{P}_\text{HZ} = (H/2\pi)^{2}$.

\subsection{Quantization in Fourier space}

The algorithm we just described involved quantization of $\phi$ first and then decomposition into Fourier modes.  However, there is a equivalent procedure that involves Fourier decomposition and then quantization.  The first step is writing
 \begin{align}
\phi(t,\x) &=  \frac{1}{\sqrt{V_{0}}}\sum_{\k}{\phi}_{\k}(t) e^{i {\k}\cdot{\x}}, \nn\\
{\phi}_\k(t) &= \frac{1}{\sqrt{V_{0}}}\int d^3x\ e^{-i\k\cdot \x} \phi(t,\x),
\end{align}
with a similar expansion for $\pi(t,\x)$.  After  a suitable redefinition of the independent modes to ensure that $\phi$ and the redefined $\phi_{\k}$'s are real, the  Hamiltonian is 
\begin{equation}\label{eq:Fourier H}
{\scr H}_{\phi} =  \sum_{\k} {\scr H}_{\k} = \sum_{\k} \left[ \frac{\pi_{\k}^2}{2a^{3}} +
\frac{k^{2}}{2a^{2}} a^{3}\phi_{\k}^2 \right],
\end{equation}
with the Poisson bracket $\{\phi_\k, \pi_{\k'}\} = \delta_{\k,\k'}$. 

This expression of the Hamiltonian (\ref{eq:Fourier H}) implies that we can view the classical system as a collection of independent oscillators with time-dependent parameters and labeled by the wavevector $\k$.  The quantum state of the field is of the form
\begin{equation}
	|\psi \rangle = \bigotimes_{\k} |\psi_{\k} \rangle,
\end{equation}
where each of the $| \psi_{\k} \rangle$ satisfy the time-dependent Schr\"odinger equation (TDSE),
\begin{equation}\label{eq:t dependent Schrodinger}
	\hat{\!\!\scr H}_{\!\!\k} |\psi_\k \rangle = i \di_{t} |\psi_\k \rangle.
\end{equation}
As described by  \cite{Mahajan:2007qc,Mahajan:2007qg}, to determine the spectrum of primordial perturbations generated during inflation one first solves the (TDSE) for the ``ground state'' $|\psi_{\k} \rangle = |0_\k \rangle$ of each mode.  Then, the power spectrum is given by:
\begin{equation}
	\mathcal{P}_{\phi}(k) = \frac{k^{3}}{2\pi^{2}}  \langle \phi_{\k}^{2} \rangle\bigg|_{k \ll Ha}, \quad \langle \phi_{\k}^{2} \rangle = \langle 0_\k | \hat{\phi}_{\k}^{2} | 0_\k \rangle.
\end{equation}
This method is useful because it reduces the full quantum field theory to problem in quantum mechanics.  This is crucial, because PQ is much easier to deal with a purely quantum mechanical setting.  

\section{Schr\"odinger quantization of an individual Fourier mode}\label{sec:SQ}

We are ultimately interested in the solution of the TDSE (\ref{eq:t dependent Schrodinger}) with the polymer representation of the Hamiltonian operator governing a single Fourier mode, but as a prelude we review how the calculation works using standard Schr\"odinger quantization (SQ).  Recall that we can represent arbitrary quantum states as wavefunctions depending on either ``position'' or ``momentum'', which in our case correspond to $\phi_{\k}$ or $\pi_{\k}$, respectively.  In the case of SQ, both choices are very similar due to the symmetric form of the simple harmonic oscillator Hamiltonian (\ref{eq:Fourier H}).  However as described in Appendix \ref{sec:polymer}, the polymer quantization of position and momentum operator are handled in quite different ways, which has the net effect of making it easier to work with momentum space wavefunctions.  Hence, we assume arbitrary quantum states are represented by functions of  $\pi_{\k}$:
\begin{equation}\label{eq:momentum wavefunction}
	\langle \pi_{\k} | \psi_{\k} \rangle = \psi(t,\pi_{\k}),
\end{equation}
where $| \pi_{\k} \rangle$ is a momentum eigenstate.  The action of $\hat{\phi}_{\k}$ and $\hat{\pi}_{\k}$ on these wavefunctions is simply:
\begin{subequations}
\begin{align}
	\langle \pi_{\k} | \hat\phi_{\k} | \psi_{\k} \rangle &  = i {\di_{\pi_{\k}}} \psi(t,\pi_{\k}), \\ \langle \pi_{\k} | \hat\pi_{\k} | \psi_{\k} \rangle & = \pi_{k} \psi(t,\pi_{\k}).
\end{align}
\end{subequations}
It is easy to confirm that this operator representation respects the commutation relation $[\hat\phi_{\k},\hat\pi_{\k}] = i$.   Using these, we find that the action of the Hamiltonian operator on an arbitrary state is:
\begin{equation}\label{eq:Schrodinger H}
	\langle \pi_{\k} | \,\, \hat{\!\!\scr H}_{\!\!\k} | \psi_{\k} \rangle = \left[ \frac{1}{2\mu}\pi_{\k}^{2} - \frac{\mu\omega^{2}}{2} \frac{\di^{2}}{\di\pi_{\k}^{2}} \right] \psi(t,\pi_{\k});
\end{equation}
where we have defined the time-dependant parameters
\begin{equation}\label{eq:mu and omega def}
	\mu = a^{3}, \quad \omega = k/a.
\end{equation}

With the representation (\ref{eq:Schrodinger H}), the time-dependant Schr\"odinger equation,
\begin{equation}\label{eq:time dep Schro}
	\hat{\!\!\scr H}_{\!\!\k} |\psi_{\k} \rangle = i \di_{t} | \psi_{\k} \rangle,
\end{equation}
reduces to a PDE for the wavefunction $\psi$.  If we now change to the conformal time $\eta$ and a dimensionless momentum $y$,
\begin{equation}\label{eq:coord transform}
	\eta = -\frac{1}{Ha}, \quad y = \frac{\pi_{\k}}{\sqrt{\mu\omega}} = -k\eta \sqrt{\frac{H^{2}}{k^{3}}} \pi_{\k},
\end{equation}
and re-scale the wavefunction as
\begin{equation}\label{eq:Psi def}
	\psi(t,\pi_{k}) = \left( \frac{H^{2}}{k^{3}} \right)^{1/4} \sqrt{-k\eta} \Psi(\eta,y) \exp \left( -i\frac{y^{2}}{2k\eta} \right),
\end{equation}
we find that
\begin{subequations}\label{eq:SHO Schrodinger}
\begin{gather}
	i \di_{\eta} \Psi(\eta,y) = \hat{\mathsf{H}} \Psi(\eta,y), \\ \hat{\mathsf{H}} = k\left(\frac{1}{2} y^{2} - \frac{1}{2} \frac{\di^{2}}{\di y^{2}}\right).
\end{gather}
\end{subequations}
We see that the new wavefunction is just a solution of the familiar TDSE for an oscillator with unit mass and frequency---and the new effective Hamiltonian $\hat{\mathsf{H}}$ is time independent.\footnote{We stress that the effective Hamiltonian $\hat{\mathsf{H}}$ should be viewed as a linear differential operator rather than a true quantum Hamiltonian; for example, it does not generate the time evolution of observables.  Having said this, we will use the terms ``ground state'', ``energy eigenstate'', ``eigenenergy'', etc. when referring to solutions of the ODE eigenvalue problem for $\hat{\mathsf{H}}$ in this subsection and below.}  Normalizable solutions of (\ref{eq:SHO Schrodinger}) satisfy the boundary conditions
\begin{equation}\label{eq:SQ BCs}
	\Psi(\eta,\pm \infty) = 0,
\end{equation}
and are well known:
\begin{equation}
	\Psi(\eta,y) = \sum_{n=0}^{\infty} c_{n} \Psi_{n}(y) e^{-i(n+1/2)k\eta}, \quad \sum_{n=0}^{\infty} |c_{n}|^{2} = 1,
\end{equation}
with
\begin{equation}\label{eq:Schrodinger eigenfunctions}
	\Psi_{n}(y) = (\sqrt{\pi} 2^{n} n!)^{-1/2}H_{n}(y) e^{-y^{2}/2}.
\end{equation}
Here, $H_{n}$ is the Hermite polynomial of order $n$.  Notice that the constants in the $\Psi$ definition ensure that if $\Psi$ is normalized with respect to integration over $y$, $\psi$ is also normalized with respect to integration over $\pi_{\k}$:
\begin{equation}
	1 = \int_{-\infty}^{\infty} dy |\Psi(\eta,y)|^{2} = \int_{-\infty}^{\infty} d\pi_{\k} |\psi(t,\pi_{k})|^{2}
\end{equation}

Having now obtained the explicit solution for the TDSE, we need to identify the ground state wavefunction.  The most natural choice is to use the ground state of $\hat{\mathsf{H}}$,
\begin{equation}
	\Psi(\eta,y) = \Psi_{0}(y) e^{-ik\eta/2} = \pi^{-1/4} e^{-(y^{2}+ik\eta)/2},
\end{equation}
since this state minimizes the expectation value of the effective Hamiltonian for all $\eta$.  Indeed, it is precisely this choice that reproduces the familiar Bunch-Davies result:
\begin{align}\nn
	\langle \phi_{\k}^{2} \rangle & = \int_{-\infty}^{\infty} \psi^{*} \left(i \frac{\di}{\di\pi_{\k}} \right)^{2}\psi \, d\pi_{\k} \\  
	\nn & = \frac{H^{2}}{k^{3}} \int_{-\infty}^{\infty} \left[y^{2} |\Psi|^{2} +  \frac{k^{2}}{H^{2}a^{2}} |\di_{y}\Psi|^{2} \right] \, dy   \\
	& = \frac{H^{2}}{2k^{3}} \left(1 + \frac{k^{2}}{H^{2}a^{2} }\right).
\end{align}
Just as in the standard calculation, we would have obtained a different answer had we imposed different conditions on the quantum state of the system.

\section{Polymer quantization of an individual Fourier mode}\label{sec:PQ}

\subsection{Formal solution of time dependent Schr\"odinger equation}\label{sec:PQ formal}

We now seek to find a ``ground state solution'' of the polymer version of the TDSE governing the amplitude of a given mode in $\k$-space (\ref{eq:t dependent Schrodinger}), from which we can calculate $\mathcal{P}_{\phi}$.  We again work with momentum wavefunctions $\psi(t,\pi_{\k})$.  Our treatment will largely follow the discussion of the polymer quantization of a particle moving in a one-dimensional potential in Appendix \ref{sec:polymer}, with a few notable exceptions.

As in the Schr\"odinger case of \S\ref{sec:SQ}, we represent the quantum state of an individual Fourier mode in a basis $|\pi_{\k}\rangle$:
\begin{equation}
	\langle \pi_{\k} | \psi_{\k} \rangle = \psi(t,\pi_{\k}).
\end{equation}
However in the polymer scenario, $|\pi_{\k}\rangle$ is not interpreted as a momentum eigenstate because the momentum operator $\hat{\pi}_{\k}$ does not exist.  But we can define an operator $\hat{U}_{\lambda}$ that will be seen to correspond to finite translations of the field amplitude:
\begin{equation}
	\langle \pi_{\k} | \hat{U}_{\lambda} | \psi_{\k} \rangle = \exp ( i\lambda a^{-3/2} \pi_{\k} )\psi(t,\pi_{\k}).
\end{equation}
This definition is similar to the one given in Appendix \ref{sec:polymer}, except for the $a^{-3/2}$ factor in the argument of the exponential.  As in Ref.\ \cite{Hossain:2009ru}, this has been included to ensure that $\hat{U}_{\lambda}$ transforms as a scalar under the dilation $\x \rightarrow \ell \x$ (\emph{cf.} Appendix \ref{sec:scaling}), which will ensure we recover the correct scaling of the effective momentum given below.  We define the operator corresponding to the field amplitude $\phi_{\k}$ in the same manner as the SQ case:
\begin{align}
	\langle \pi_{\k} | \hat\phi_{\k} | \psi_{\k} \rangle &  = i {\di_{\pi_{\k}}} \psi(t,\pi_{\k}).
\end{align}
It is easy to confirm the following commutation relation holds:
\begin{equation}
	[\hat\varphi_{\k},\hat{U}_{\lambda}] = -\lambda \hat{U}_{\lambda},
\end{equation}
where we have defined the smeared Fourier amplitude operator
\begin{equation}
	\hat\varphi_{\k} = a^{3/2} \hat{\phi}_{\k},
\end{equation}
which transforms as a scalar under $\x \rightarrow \ell\x$.  This allows us to further interpret $\hat{U}_{\lambda}$:  Suppose $|\varphi_{\k}\rangle$ is an eigenstate of $\hat\varphi_{\k}$:
\begin{equation}
	\hat{\varphi}_{\k} |\varphi_{\k}\rangle = \varphi_{\k} |\varphi_{\k}\rangle.
\end{equation}
Then $\hat{U}_{\lambda}|\varphi_{\k}\rangle$ will be an eigenstate of $\hat\varphi_{\k}$ with eigenvalue $\varphi_{\k} - \lambda$:
\begin{align}
	\nonumber \hat{\varphi}_{\k} \left(\hat{U}_{\lambda} |\varphi_{\k}\rangle\right) & = \left( \hat{U}_{\lambda} \hat{\varphi}_{\k} - \lambda  \hat{U}_{\lambda}  \right) |\varphi_{\k}\rangle \\ & =  \left( \varphi_{\k} - \lambda \right) \left(\hat{U}_{\lambda} |\varphi_{\k}\rangle\right).
\end{align}
Hence, $\hat{U}_{\lambda}$ has the effect of inducing translations of magnitude $\lambda$ in the smeared Fourier amplitude $\varphi_{\k}$.

Once we have a representation of $\hat{U}_{\lambda}$, we realize the momentum contained in the Hamiltonian operator $\,\,\hat{\!\!\scr H}_{\!\!\k}$ as a finite difference operator:\footnote{It is important to note that this realization of the momentum operator in polymer quantum mechanics is not unique; i.e., it represents a new type of quantization ambiguity in addition to ones already present in conventional Schr\"odinger quantization.  Stated another way: our quantization scheme is defined by our choice of fundamental operators $\hat{\varphi}_{\k}$ and $\hat{U}_{\lambda}$ \emph{as well as} our specification of the momentum operator (\ref{eq:p def}).  Whether or not other finite difference representations of the momentum yield different physical results is an open question.\label{foot:ambiguity}}
\begin{equation}\label{eq:p def}
	\hat\pi_{\k} \mapsto \hat{\pi}^{\star}_{\k} =  \frac{a^{3/2}}{2i\lambda_{\star}}(\hat{U}_{\lambda_{\star}}-\hat{U}_{\lambda_{\star}}^{\dag}),
\end{equation} 
where $\lambda_{\star} \equiv M_{\star}^{-1/2}$ is a fixed parameter with dimensions of $(\text{mass})^{-1/2}$.   Since $\hat{U}_{\lambda_{\star}}$ transforms as a scalar, we see that $\hat{\pi}^{\star}_{\k}$ transforms like $\pi_{\k}$ under dilations.  Furthermore, we recover the SQ momentum operator in the appropriate limit:
\begin{equation}
	\lim_{\lambda_{\star}\rightarrow 0} \langle \pi_{\k} | \hat{\pi}^{\star}_{\k} | \psi_{\k} \rangle = \lim_{a\rightarrow \infty} \langle \pi_{\k} | \hat{\pi}^{\star}_{\k} | \psi_{\k} \rangle = \pi_{\k} \psi(t,\pi_{\k}).
\end{equation}
It is fairly easy to confirm that this gives the action of $\,\hat{\!\! \scr{H}}_{\!\!\k}$ on an arbitrary state as
\begin{equation}\label{eq:polymer H}
	\langle \pi_{\k} | \,\,\hat{\!\! \scr H}_{\!\!\k} | \psi_{\k} \rangle = \left[ \frac{\sin^{2} (\Lambda\pi_{\k})}{2\mu\Lambda^{2}} - \frac{\mu\omega^{2}}{2} \frac{\di^{2}}{\di\pi_{\k}^{2}} \right] \psi(t,\pi_{\k}),
\end{equation}
where $\mu$ and $\omega$ are defined as above (\ref{eq:mu and omega def}), while
\begin{equation}
	\Lambda = {\lambda}{a^{-3/2}}.
\end{equation}
In the $\Lambda\pi_{\k} \rightarrow 0$ limit, we see that (\ref{eq:polymer H}) reduces to the SQ expression (\ref{eq:Schrodinger H}).  As described in detail in Appendix \ref{sec:polymer}, if we restrict ourselves to one super-selected sector of the polymer Hilbert space the appropriate inner product between states is:
\begin{equation}
	\langle \xi_{\k} | \psi_{\k} \rangle = \int_{-\pi /2\Lambda}^{\pi /2\Lambda} \xi^{*}(t,\pi_{\k}) \psi(t,\pi_{\k}) \, d\pi_{\k}.
\end{equation}
Furthermore, wavefunctions will satisfy
\begin{equation}\label{eq:momentum shift}
	\psi(t,-\pi/2\Lambda) = \exp(i\pi\phi_{\k}^{0}/\Lambda) \psi(t,\pi/2\Lambda),  
\end{equation}
where $\phi_{\k}^{0}/\Lambda \in [0,2)$ is a constant (this is the lattice offset of the super-selected sector, as discussed in Appendix \ref{sec:polymer}).

As in the SQ case, our task is to now solve the TDSE with the polymer Hamiltonian.  We again change coordinates and re-scale the wavefunction as in Eqs.\ (\ref{eq:coord transform}) and (\ref{eq:Psi def}) to obtain
\begin{subequations}\label{eq:SHO polymer}
\begin{gather}
	i \frac{\di}{\di\eta} \Psi(\eta,y) =  \hat{\mathsf{H}} \Psi(\eta,y), \\ \hat{\mathsf{H}} =  k\left[ \frac{\sin^{2}(g^{1/2}y)}{2g} - \frac{1}{2} \frac{\di^{2}}{\di y^{2}} \right],
\end{gather}
\end{subequations}
where the \emph{polymer coupling} is defined by
\begin{equation}\label{eq:g def}
	g = g(\eta) = \mu\omega\Lambda^{2} = \frac{k}{M_{\star} a} = -(k\eta) \left( \frac{H}{M_{\star}} \right).
\end{equation} 
Unlike the SQ case, the effective Hamiltonian $\hat{\mathsf{H}}$ is a function of time via its dependence on the coupling $g$.  At late times $\eta \rightarrow 0$ and $g \rightarrow 0$, which gives $\hat{\mathsf{H}} \rightarrow \frac{1}{2}(y^{2}-\di_{y}^{2})$; i.e.\ we recover the SQ Hamiltonian.  Note that under this change of coordinates and scaling of the wavefunction, the inner product becomes
\begin{equation}\label{eq:inner product}
	\langle \xi_{\k} | \psi_{\k} \rangle = \int\limits_{y\in I}  \Xi^{*}(\eta,y) \Psi(\eta,y) \, dy,
\end{equation}
where $\psi(t,\pi_{\k})$ and $\xi(t,\pi_{\k})$ are the images of $\Psi(\eta,y)$ and $\Xi(\eta,y)$ under the wavefunction transformation defined by (\ref{eq:Psi def}), respectively, and
\begin{equation}
	I = [-\pi/2\sqrt{g},\pi/2\sqrt{g}].
\end{equation}

We now seek a solution of (\ref{eq:SHO polymer}) in terms of the eigenfunctions of the time-dependent Hamiltonian $\hat{\mathsf{H}}$.  To fix these eigenfunctions, we need to specify boundary conditions on $\Psi(\eta,y)$.  The relationship (\ref{eq:momentum shift}) and the wavefunction transformation (\ref{eq:Psi def}) imply\footnote{This could have also been deduced by demanding that the probability amplitude $|\Psi|^{2}$ share the same periodicity as the effective potential appearing in (\ref{eq:SHO polymer}).  Or, by demanding that $\hat{\phi}_{\k}$ be self-adjoint under the inner product (\ref{eq:inner product}).}
\begin{equation}\label{eq:periodicity}
	|\Psi(\eta,-\pi/2\sqrt{g})|^{2} = |\Psi(\eta,\pi/2\sqrt{g})|^{2}.
\end{equation}
Now, we impose that the evolution of an arbitrary state $|\psi_{\k}\rangle$ be unitary; that is,
\begin{equation}
	\frac{d}{d\eta} \langle \psi_{\k} | \psi_{\k} \rangle = 0.
\end{equation}
Carrying out the differentiation by making use of (\ref{eq:SHO polymer}), (\ref{eq:inner product}), and (\ref{eq:periodicity}) we obtain:
\begin{equation}\label{eq:BCs}
	\Psi(\eta,\pm \pi/2\sqrt{g}) = 0;
\end{equation}
that is, $\Psi$ must satisfy Dirichlet boundary conditions.  Note that this is consistent with the boundary condition (\ref{eq:polymer BC}), which enforces that the spectrum of the Hamiltonian in polymer quantum mechanics is independent of lattice offset.  Finally, note that (\ref{eq:BCs}) recovers the SQ boundary conditions (\ref{eq:SQ BCs}) in the $g\rightarrow 0$ limit.

\begin{figure}
\includegraphics[width=\columnwidth]{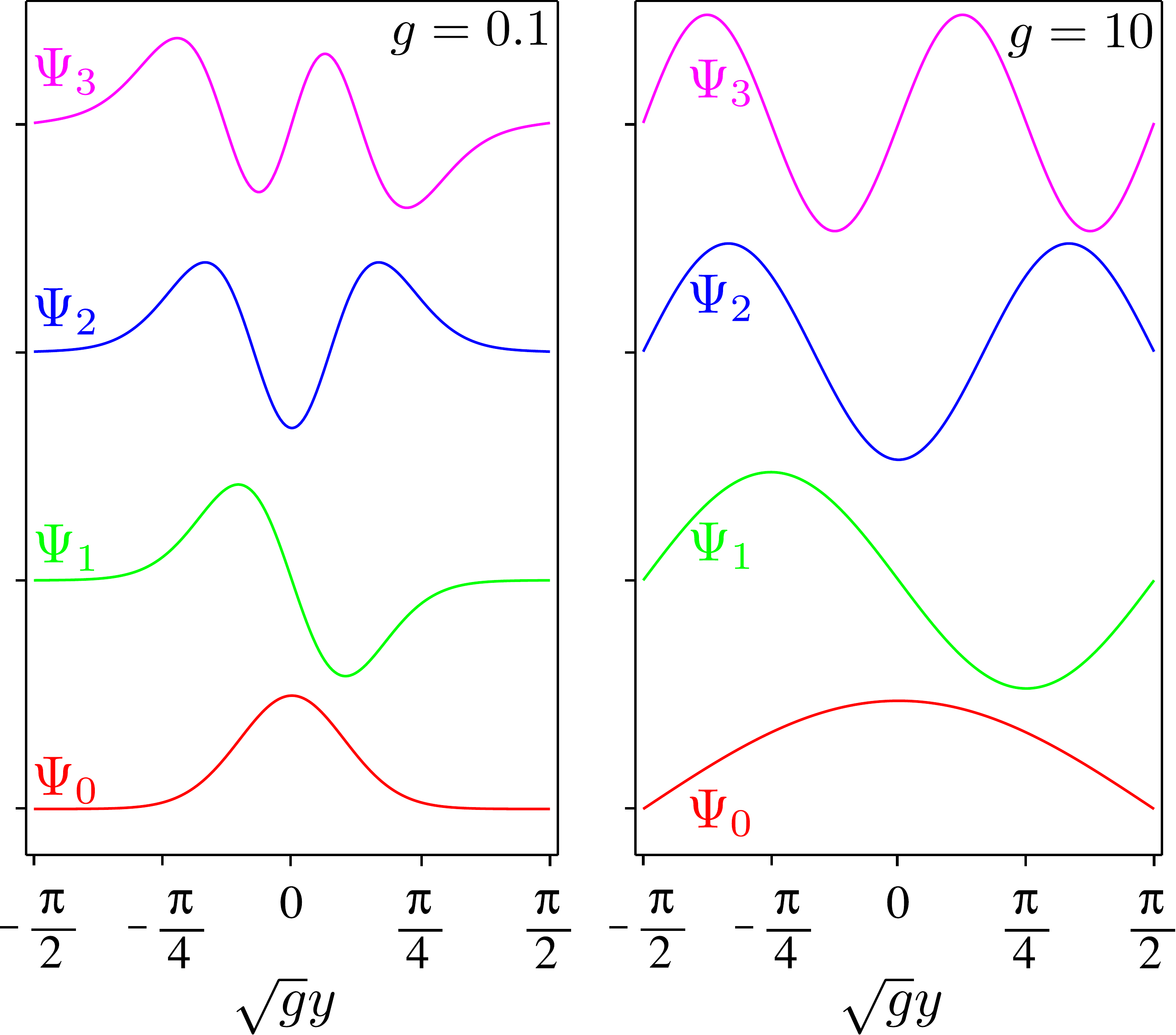}
\caption{Instantaneous energy eigenfunctions of the effective polymer Hamiltonian.  At moderately small polymer coupling (\emph{left}), these resemble the energy eigenstates of the simple harmonic oscillator, while at larger coupling (\emph{right}) they reduce to trigonometric functions.}\label{fig:eigenfunctions}
\end{figure}

\begin{figure}
\includegraphics[width=\columnwidth]{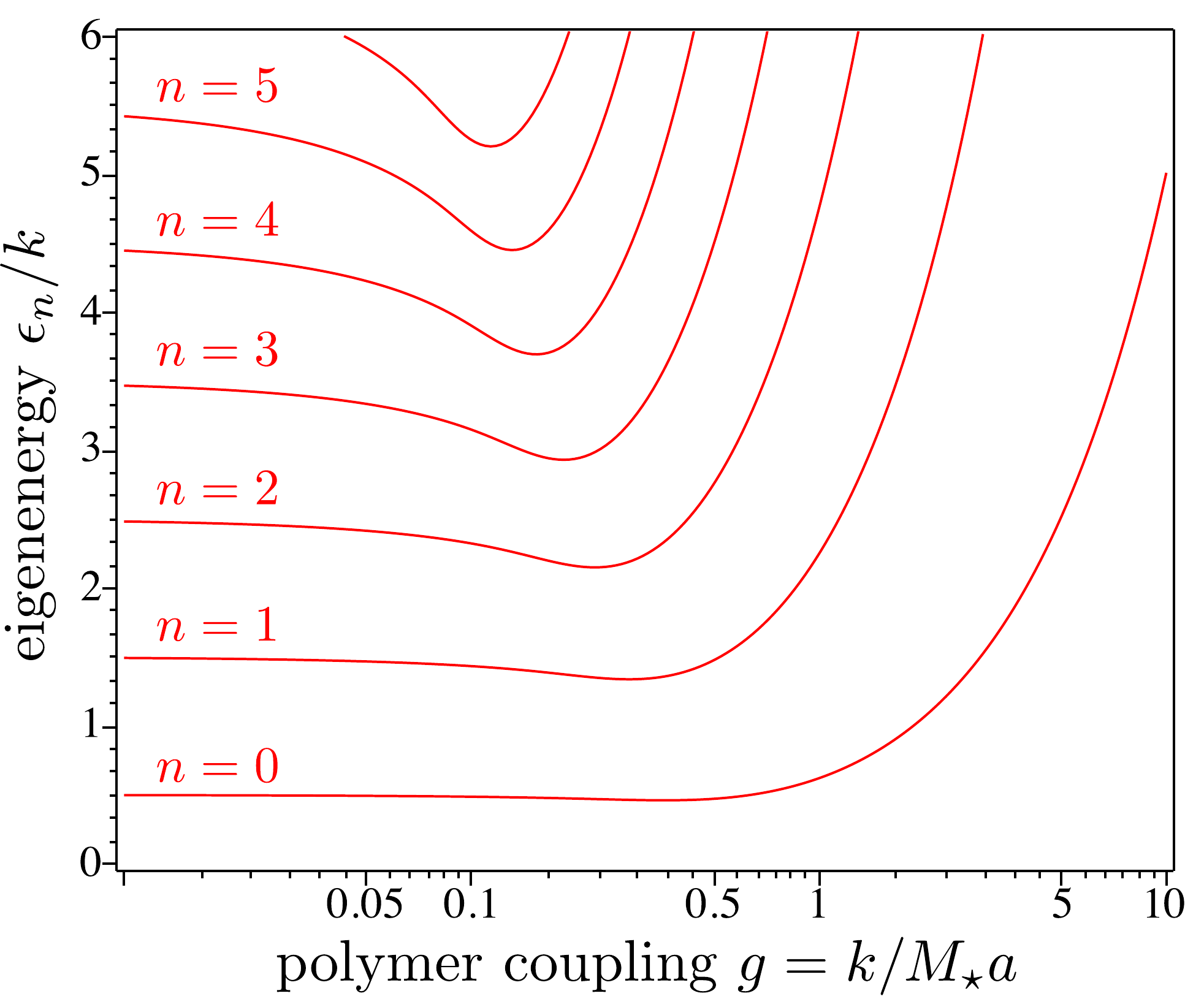}
\caption{Energy eigenvalues as a function of $g$ of the effective polymer Hamiltonian}\label{fig:eigenenergies}
\end{figure}

The solution of the energy eigenvalue problem
\begin{equation}
	\hat{\mathsf{H}}(\eta) \Psi_{n}(\eta,y) = \epsilon_{n}(\eta) \Psi_{n}(\eta,y),
\end{equation}
subject to the boundary conditions (\ref{eq:BCs}) is given explicitly in terms of Mathieu (elliptic sine) functions:
\begin{align}
	\Psi_{n}(g,y) &=  \sqrt{\frac{2}{\pi}}g^{1/4} \mathrm{se}_{n+1}\left( \frac{1}{4g^2},\sqrt{g}y+\frac{\pi}{2} \right),
\end{align}
with eigenvalues given by the Mathieu characteristic value functions:
\begin{align}
  \epsilon_{n} =  \frac{k}{4} \left[ {2g B_{n+1}\left( \frac{1}{4g^{2}} \right)+\frac{1}{g}} \right],
\end{align}
for $n = 0,1,2\ldots$  These form an instantaneous orthonormal basis for arbitrary functions on $I$ satisfying (\ref{eq:BCs}):
\begin{equation}
	\langle n | m \rangle \equiv \int\limits_{y \in I} \Psi^{*}_{n} \Psi_{m}  \, dy = \delta_{nm}.
\end{equation}
Also, eigenfunctions with even $n$ have even parity and those with odd $n$ have odd parity: $\Psi_{n}(-y) = (-1)^{n}\Psi_{n}(y)$.  We can use the asymptotic expansions of the Mathieu functions \cite{2001JPhA...34.3541F} to deduce
\begin{equation}\label{eq:eigenfunction approx}
	\Psi_{n}(g,y) \rightarrow \begin{cases} \displaystyle \frac{1}{(\sqrt{\pi} 2^{n} n!)^{1/2}} H_{n}(y) e^{-y^{2}/2} , & \!\! \displaystyle g \ll G_{n}, \\ \displaystyle
	\sqrt{ \frac{2}{\pi} } g^{1/4}\sin\left[ (n+1)\left(\sqrt{g}y+\frac{\pi}{2}\right)\right], & \!\! \displaystyle g \gg G_{n}, \end{cases}
\end{equation}
where $G_{n}= 1/(n+1/2)$; i.e., we recover the Schr\"odinger energy eigenfunctions for small polymer coupling and simple trigonometric function for large coupling.  Asymptotic expansions of the Mathieu characteristic value function $B_{n}$ yield the following approximations for the eigenenergies:
\begin{equation}\label{eq:eigenenergy approx}
	\epsilon_{n} \rightarrow k \begin{cases} \displaystyle n+\tfrac{1}{2} , & g \ll G_{n}, \\ \displaystyle
	\tfrac{1}{2}(n+1)^{2} g, & g \gg G_{n}. \end{cases}
\end{equation}
Plots of the energy eigenfunctions and eigenvalues are presented in figures \ref{fig:eigenfunctions} and \ref{fig:eigenenergies}, respectively.

Since our effective Hamiltonian is time dependent, arbitrary solutions of (\ref{eq:SHO polymer}) can be constructed from Hamiltonian eigenstates if we allow the expansion coefficients to depend on $\eta$:
\begin{equation}\label{eq:Psi decomp}
	|\Psi\rangle = \sum_{n=0}^{\infty} c_{n}(\eta) e^{i\theta_{n}(\eta)} |n \rangle, 
\end{equation}
where $\theta_{n}(\eta)$ satisfies
\begin{equation}
	\dot{\theta}_{n} = -\epsilon_{n}, \quad \theta_{n}(\eta_{0}) = 0,
\end{equation}
Here $\eta_{0}$ is some initial time, and we use an overdot to indicate derivatives with respect to $\eta$. Some straightforward algebra reveals that the expansion coefficients satisfy an autonomous set of linear ODEs:
\begin{equation}\label{eq:c ODEs}
	\dot{c}_{n} = -c_{n} \langle n | \dot{n} \rangle - \sum_{m \ne n} c_{m} \frac{\langle n | \dot{\hat{\mathsf{H}}} | m \rangle}{\epsilon_{m}-\epsilon_{n}} e^{i(\theta_{m}-\theta_{n})}.
\end{equation}
The normalization condition $\langle n | n \rangle = 1$ in addition to the fact that the energy eigenfunctions are real imply that $\langle n | \dot{n} \rangle = 0$.  Since the polymer coupling $g$ is proportional to the conformal time $\eta$ (\ref{eq:g def}), we can recast this as a first order matrix ODE:
\begin{equation}\label{eq:c evolution}
	\frac{d}{dg} \mathbf{c} = \mathbf{A} \mathbf{c}, \quad \mathbf{c} = \left[ \begin{array}{c} c_{0} \\ c_{1} \\ \vdots \end{array} \right] , \quad \mathbf{A} = \left[ \begin{array}{ccc} a_{00} & a_{01} & \cdots \\ a_{10} & a_{11} &  \\ \vdots & & \ddots \end{array} \right], 
\end{equation}
where
\begin{multline}\label{eq:a matrix elements}
	a_{nm}  = -\frac{k}{\epsilon_{m}-\epsilon_{n}} \exp\left[ i\frac{\M}{H} \int_{g_{0}}^{g} \left( \frac{\epsilon_{m}-\epsilon_{n}}{k} \right) d\tilde{g} \right] \\ \times \int\limits_{y \in I}  \Psi_{m} \frac{d}{dg} \left[ \frac{\sin^{2}(g^{1/2}y)}{2g}  \right] \Psi_{n} dy,
\end{multline}
for $n \ne m$ and $a_{nn} = 0$.  Here, $g_{0}$ is that value of the polymer coupling at time $\eta_{0}$.  Note that the matrix $\mathbf{A}$ is anti-Hermitian ($\mathbf{A}^{\dag} = - \mathbf{A}$) so the norm of $\mathbf{c}$ is conserved:
\begin{equation}
	\frac{d}{dg} (\mathbf{c}^{\dag} \mathbf{c}) = 0.
\end{equation}
Also note that $a_{ij}$ will be non-zero only if $\Psi_{n}$ and $\Psi_{m}$ have the same parity; i.e., $a_{i,i+2k+1} = 0$ for $k = 0, \pm 1, \pm 2, \ldots$

\subsection{Initial conditions and the final spectrum of fluctuations}

A unique solution of the polymer TDSE will be characterized by the specification of initial conditions for the expansion coefficients at some time $\eta_{0}$.  The question is: what choice of initial conditions could reasonably be associated with the vacuum state of a given Fourier mode?  In the case of SQ, the answer was straightforward because the effective Hamiltonian $\hat{\mathsf{H}}$ was independent of $\eta$, implying that if we prepared a given mode in its ground state at some initial time, it would stay in its ground state indefinitely.  That is, the quantum evolution was perfectly adiabatic.  This is not true in the PQ case: (\ref{eq:c ODEs}) tells us that there is non-trivial mode mixing.  That is, if we prepare the system in the ground state at some early time, it will not be in the ground state at the end of inflation.

The situation is akin to the trans-Planckian problem of inflationary cosmology considered by Martin and Brandenberger \cite{Martin:2000xs}.  In that work, the classical equations of motion of the inflaton were modified on small scales in a attempt to account for quantum gravity effects.  The net result was that one could not unambiguously identify a vacuum state in the asymptotic past due to particle creation induced by the modified wave equation, which is precisely analogous to the mode mixing induced by polymer effects at early times in the current scenario.  Hence, we will adopt the same prescription for initial conditions as employed in \cite{Martin:2000xs}:  We assume that the field is in the instantaneous ground state at the \emph{beginning} of inflation.  That is,
\begin{equation}\label{eq:c IC}
	\mathbf{c}(g_{0}) = \left[ \begin{array}{ccc} 1 & 0 & \cdots \end{array} \right]^\dag, \quad g_{0} = g(\eta_{0}), \quad  \eta_{0} = -\frac{1}{Ha_{0}}.
\end{equation}
Here, $a_{0}$ is the scale factor at the beginning of inflation.  Notice that the evolution equation for the coefficients (\ref{eq:c evolution}) is in terms of polymer coupling instead of the conformal time, so it useful to know that the initial value of $g$ for a given mode
\begin{equation}
	g(\eta_{0}) = g_{0} = \frac{k}{k_{\star}}, \quad k_{\star} = M_{\star} a_{0} = - \frac{M_{\star}}{H\eta_{0}}.
\end{equation}
Here, the pivot scale $k_{\star}$ is the wavenumber of a mode that has physical wavelength $\M^{-1}$ at the beginning of inflation.  As derived in Appendix \ref{sec:k star derivation}, the numeric value of $k_{\star}$ is
\begin{equation}\label{eq:k star value}
	k_{\star} \sim \frac{6\times10^{-6}}{\text{Mpc}} \left( \frac{M_{\star}}{H} \right)  \left( \frac{E_{\text{inf}}}{10^{16}\,\text{GeV}} \right) \left( \frac{e^{65}}{e^{N}} \right)  \left( \frac{100}{\mathcal{G}} \right)^{1/12},
\end{equation}
where $E_{\text{inf}} = \rho_{\inf}^{1/4}$ is the energy scale of inflation,  $N = \ln (a_{\text{end}}/a_{0})$ is the number of $e$-folds of inflation, and $\mathcal{G}$ is the effective number of relativistic species at the end of inflation.
\begin{figure}
\includegraphics[width=\columnwidth]{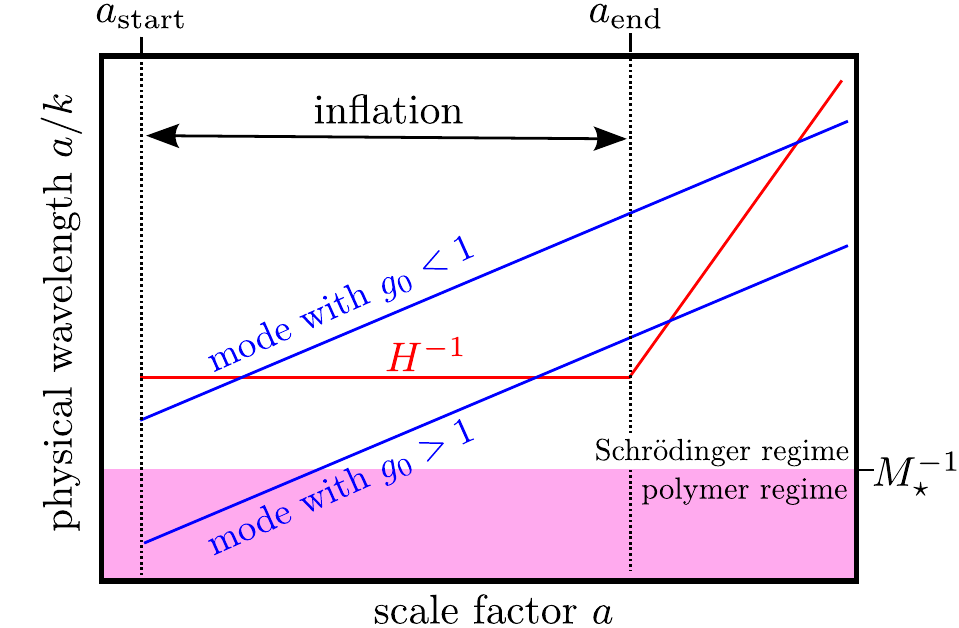}
\caption{Scale factor evolution of the physical wavelength of modes with large or small $g_{0}$ relative to the polymer $\M^{-1}$ and Hubble $H^{-1}$ scales}
\end{figure}

Having now specified the quantum state at beginning of inflation, we note that observable quantities are directly derived from the quantum state of the end of inflation.   We will restrict our discussion to modes with physical wavelengths much less than $\M^{-1}$ at the end of inflation; i.e, modes with
\begin{equation}
	g_{\text{end}} = e^{-N} g_{0} \ll 1.
\end{equation} 
Hence, the final quantum state of the system will be given by
\begin{equation}
	\mathbf{c}^{\text{end}} \simeq \mathbf{c}(0). 
\end{equation}
We are also interested in modes that are well into the superhorizon regime at the end of inflation; i.e., modes with
\begin{equation}
	\frac{k}{Ha_{\text{end}}} = \frac{\M}{H} e^{-N}g_{0} \ll 1.
\end{equation}
Note that assuming $H \lesssim \M$ implies that all superhorizon modes at the end of inflation will automatically have $g_\text{end} \ll 1$.
Using the definition (\ref{eq:Psi def}), we see that power spectrum of such modes is
\begin{equation}\label{eq:polymer power spectrum}
	\mathcal{P}_{\phi}(k) = \frac{k^{3}}{2\pi^{2}} \langle \phi_{\k}^{2} \rangle\bigg|_{k \ll aH} \approx \left( \frac{H}{2\pi} \right)^{2} \langle \Psi | 2y^{2} | \Psi \rangle \bigg|_{k \ll aH}, 
\end{equation}	 
where we have assumed $k/Ha \ll 1$ to obtain the last expression.  The state vector in this expression is given by the $\eta \rightarrow 0$ limit of (\ref{eq:Psi decomp}):
\begin{equation}\label{eq:polymer power spectrum 2}	  
	 |\Psi \rangle\bigg|_{k \ll aH} = \sum_{n = 0}^{\infty} c_{n}^{\text{end}} | n \rangle_{\text{SQ}},
\end{equation}
where $|n\rangle_{\text{SQ}}$ are the energy eigenstates of the ordinary Schr\"odinger quantized simple harmonic oscillator.

To summarize, in order to calculate the power associated with a mode with a given value of $g_{0} = k/k_{\star}$ at the end of inflation, we must solve the matrix equation (\ref{eq:c evolution}) subject to the initial condition (\ref{eq:c IC}) for the final values of the expansion coefficients $\mathbf{c}(0)$.  Then, these expansion coefficients can be inserted into the formula (\ref{eq:polymer power spectrum}) to obtain $\mathcal{P}_{\phi}(k)$.

\subsection{Perturbative solutions}\label{sec:perturbative}

\subsubsection{Small coupling: $g \ll 1$}

We can solve for the expansion coefficients in the matrix ODE (\ref{eq:c evolution}) using standard perturbation theory in the small coupling regime.  This approximation will be valid provided that the mode in question has $g_{0} \ll 1$. We first expand the matrix $\mathbf{A}$ in a power series:
\begin{equation}
	\mathbf{A} = \mathbf{A}^{(0)} + \mathbf{A}^{(1)}g + \cdots
\end{equation}
and define the following perturbative expansion of $\mathbf{c}$:
\begin{equation}\label{eq:zeroth order}
	\mathbf{c} = \mathbf{c}^{(0)} + \mathbf{c}^{(1)} + \cdots, \quad \mathbf{c}^{(0)} = \left[ \begin{array}{ccc} 1 & 0 & \cdots \end{array} \right]^\dag.
\end{equation}
Then, $\mathbf{c}^{(1)}$ will satisfy the ODE
\begin{equation}\label{eq:perturbative ODE}
         \frac{d}{dg} \mathbf{c}^{(1)}  = \mathbf{A}^{(0)}  \mathbf{c}^{(0)} , \quad \mathbf{c}^{(1)} (g_{0}) = 0.
\end{equation}
For $n \ne m$, the elements of the $\mathbf{A}_{0}$ matrix are given by
\begin{equation}
	a^{(0)}_{nm} = \frac{1}{6(m-n)} \int_{-\infty}^{\infty} \Psi_{m} y^{4} \Psi_{n} \, dy = \frac{{}_{\text{SQ}}\langle n|y^{4}|m\rangle_{\text{SQ}}}{6(m-n)},
\end{equation}
where the eigenfunctions correspond to the ordinary Schr\"odinger simple harmonic oscillator (\ref{eq:Schrodinger eigenfunctions}).  The matrix element is only nonzero if the quantum numbers differ by 0, 2 or 4, which implies that the only non-zero elements of $\mathbf{c}^{(1)}$ are $c^{(1)}_{2}$ and $c^{(1)}_{4}$.    Evaluating the expansion coefficients in the $g \rightarrow 0$ limit, we obtain the final quantum state of the system to be
\begin{equation}
	|\Psi \rangle = |0\rangle_{\text{SQ}} + \frac{1}{8} g_{0} \left( \sqrt{2} |2\rangle_{\text{SQ}} + \frac{1}{\sqrt{6}} |4\rangle_{\text{SQ}} \right) + \mathcal{O}(g_{0}^{2}),
\end{equation}
which yields the large scale power spectrum
\begin{equation}\label{eq:large scale polymer power spectrum}
	\mathcal{P}_{\phi}(k) = \left( \frac{H}{2\pi} \right)^{2} \left[1 + \frac{1}{2} \frac{k}{k_{\star}} + \mathcal{O}\left( \frac{k^{2}}{k_{\star}^{2}} \right) \right], 
\end{equation}	 
where we have made use of the parameterization $g_{0} = k/k_{\star}$.

\subsubsection{Large coupling: $g \gg 1$}\label{sec:large g}

It is also relatively straightforward to solve the matrix ODE (\ref{eq:c evolution}) in the large coupling regime $g \gg 1$.  The method is similar to the small coupling procedure described above, but now the leading order contribution to the coefficient matrix $\mathbf{A}^{(0)}$ is obtained by using the large $g$ expansions of the eigenfunctions (\ref{eq:eigenfunction approx}) and eigenenergies (\ref{eq:eigenenergy approx}) in the matrix elements (\ref{eq:a matrix elements}).  Then, assuming the same form of the zeroth order solution as before (\ref{eq:zeroth order}), the first order expansion coefficients obey
\begin{subequations}\label{eq:large g EOMs}
\begin{equation}
	\frac{dc_{2}^{(1)}}{dg} = -\frac{5 \exp \left[  2i\alpha (g_{0}^{2}-g^{2}) \right]}{384 g^{3}},
\end{equation}
and
\begin{equation}
	\frac{dc_{n}^{(1)}}{dg} = -\frac{4(n+1)\exp \left[ \tfrac{1}{4} {i\alpha n(n+2)(g_{0}^{2}-g^{2})} \right]}{n^{2}(n+2)^{2}(n+4)(n-2)g^{3}},
\end{equation}
\end{subequations}
for $n = 4, 6, 8 \ldots$ (with all the coefficients with $n$ odd identically equal to zero).  Here, we have defined $\alpha = \M/H$.  As before, we assume that the mode is in the ground state at the beginning of inflation $c_{n}^{(1)}(g_{0})=0$. These ODEs are simple to integrate in terms of Gamma functions, but tend to result in long expressions that we do not reproduce here.  We will compare the results of this perturbative analysis to numeric simulations in the next subsection.

\subsection{Numerical solutions}\label{sec:numeric solutions}

Note that the perturbative results of the previous subsection can only be used to find final quantum state of modes with $g_{0} \ll 1$:  To obtain the superhorizon behaviour of modes with $g_{0} \gtrsim 1$ we must solve (\ref{eq:c evolution}) in the transition regime where the coupling is neither small nor large.  To do so, we must solve the matrix ODE numerically, which involves truncating the infinite dimensional system.  We will neglect all eigenfunctions with $n > \nmax$, which makes $\textbf{c}$ into an $(\nmax + 1)$-dimensional vector and $\mathbf{A}$ into an $(\nmax + 1)$-dimensional square matrix.  Note that $\mathbf{A}$ is still anti-Hermitian after truncation.  

It is computationally convenient to transform to a new time coordinate
\begin{equation}\label{eq:transformed EOM}
	\tau = \tau(g), \quad \frac{d\mathbf{c}}{d\tau} = \mathbf{B} \mathbf{c}, \quad \mathbf{B} \equiv  \frac{dg}{d\tau} \mathbf{A},  
\end{equation}
and introduce an evenly spaced $\tau$-lattice
\begin{equation}
	\tau_{j} = \tau_{0} - jh, \quad \tau_{0} = \tau(g_{0}).
\end{equation}
Here, $h$ is the timestep associated with our numerical scheme.  Our particular choice for the new time coordinate is
\begin{equation}
	\tau(g) = 2 \ln g + g^{2}.
\end{equation}
For small $g$, $\tau$ will be proportional to the cosmological proper time $t$.  On the other hand for large coupling we will have $\tau \approx g^{2}$, which is consistent with the time dependence of the large $g$ perturbative solutions discussed in \S\ref{sec:large g}.

We write the values $\mathbf{c}$ and $\mathbf{B}$ at a given lattice point as
\begin{equation}
	\mathbf{c}_{j} = \mathbf{c}(g(\tau_{j})), \quad \mathbf{B}_{j} = \mathbf{B}(g(\tau_{j})).
\end{equation}
A forward-Euler numerical stencil for the solution of (\ref{eq:transformed EOM}) is
\begin{equation}
	\mathbf{c}_{j+1} = \mathbf{c}_{j} - h \mathbf{B}_{j} \mathbf{c}_{j} + \mathcal{O}(h^{2}),
\end{equation}
while a backward-Euler stencil is
\begin{equation}
	\mathbf{c}_{j+1} = \mathbf{c}_{j} - h \mathbf{B}_{j+1} \mathbf{c}_{j+1} + \mathcal{O}(h^{2}).
\end{equation}
Taking the average of the forward and backward stencils and making use of $\mathbf{B}_{j+1} = \mathbf{B}_{j} + \mathcal{O}(h)$ gives
\begin{equation}\label{eq:linear system}
	(\mathbf{I} + \tfrac{1}{2} h \mathbf{B}_{j} ) \mathbf{c}_{j+1} = (\mathbf{I} - \tfrac{1}{2} h \mathbf{B}_{j} ) \mathbf{c}_{j} + \mathcal{O}(h^{2}),
\end{equation}
where $\mathbf{I}$ is the identity matrix.  Dropping the error term gives our numerical stencil:
\begin{equation}
	\mathbf{c}_{j+1} = \mathbf{U}_{j} \mathbf{c}_{j}, \quad \mathbf{U}_{j} = (\mathbf{I} + \tfrac{1}{2} h \mathbf{B}_{j} ) ^{-1}(\mathbf{I} - \tfrac{1}{2} h  \mathbf{B}_{j} ).
\end{equation}
The advantage of this numerical scheme is that the evolution operator is automatically unitary $\mathbf{U}^{\dag}_{j} \mathbf{U}_{j} = \mathbf{I}$ since $\mathbf{A}_{j}$ (and hence $\mathbf{B}_{j}$) is anti-Hermitian; hence the norm of $\mathbf{c}$ is preserved:
\begin{equation}
	\mathbf{c}_{j+1}^{\dag} \mathbf{c}_{j+1} = \mathbf{c}_{j}^{\dag} \mathbf{c}_{j} .
\end{equation}
This also implies the scheme is unconditionally stable.  The disadvantages of the scheme are that the global error is linear in the stepsize $h$, and one has to solve the linear system (\ref{eq:linear system}) for $\mathbf{c}_{j+1}$ at each timestep.  For these reasons, the method is relatively computationally expensive to implement.  On balance, we find that the unitarity and stability of the scheme are worth the additional numerical overhead.

A key element of our numerical analysis involves the efficient computation of the $a_{nm}$ matrix elements (\ref{eq:a matrix elements}).  The Mathieu functions involved in the eigenfunctions $\Psi_{n}$ are notoriously expensive to calculate numerically, so we employ the following strategy:  For $g < 10^{-2}$ or $g > 10^{2}$, we calculate $a_{nm}$ using series expansions  \cite{2001JPhA...34.3541F} of the integral
\begin{equation}
	 \int\limits_{y \in I}  \Psi_{m} \frac{d}{dg} \left[ \frac{\sin^{2}(g^{1/2}y)}{2g}  \right] \Psi_{n} dy.
\end{equation}
For $10^{-2} < g < 10^{2}$, we numerically calculate the integral at $\sim$ 30 sample points with equal logarithmic spacing and then use cubic spline interpolation to deduce the integral at other $g$ values.  We use a similar combination of series expansions and spline interpolation to efficiently calculate the energy eigenvalues $\epsilon_{n}$.

\begin{figure}
\includegraphics[width=\columnwidth]{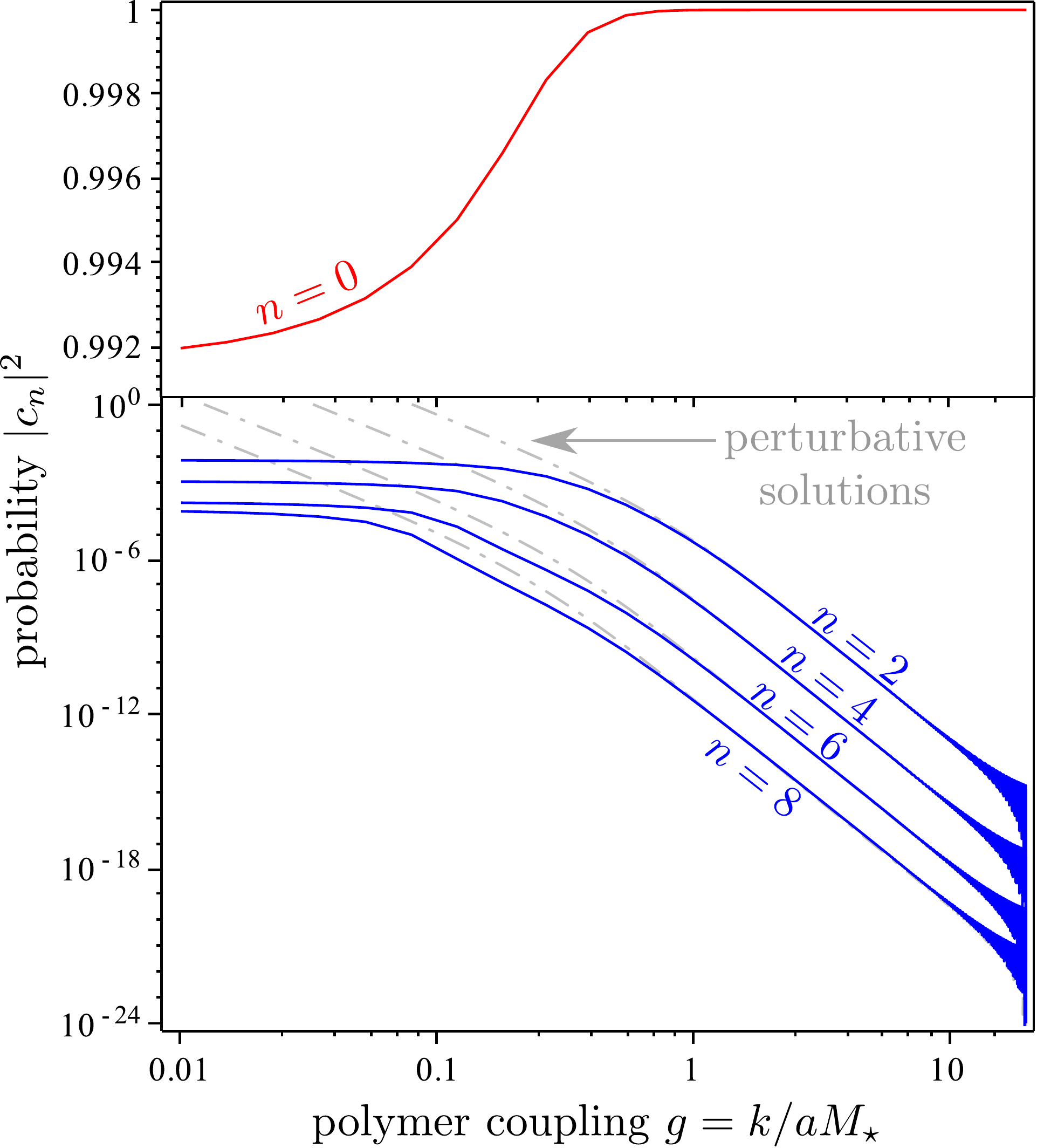}
\caption{Probability $|c_{n}|^{2}$ of finding a given Fourier mode in the $n^{\text{th}}$ energy eigenstate as a function of the polymer coupling $g$.  In this example, the mode was assumed to be in the vacuum state at an initial time characterized by $g_{0} = 20$; i.e.\ $k/a_{0}M_{\star} = 20$.   (Note that time runs right to left in this plot.)  We have selected $M_{\star}/H = 1$.  The solid lines are the results of a numerical simulation which retains the first nine energy eigenstates ($n = 0\ldots 8$) while the dash-dot lines are perturbative solutions obtained in Sec.\ \ref{sec:perturbative}.}\label{fig:numeric example}
\end{figure}

In figure \ref{fig:numeric example}, we give an example of the output of our numerical code.  We note that the numeric and perturbative solutions closely match for $g \gg 1$.  We also see that the expansion coefficients become constant in the $g \ll 1$ limit.  The transition between the two asymptotic behaviours occurs for $0.1 \lesssim g \lesssim 1$.

\section{The polymer power spectrum}\label{sec:power spectrum}

\begin{figure*}
\begin{center}
\includegraphics[width=0.7\textwidth]{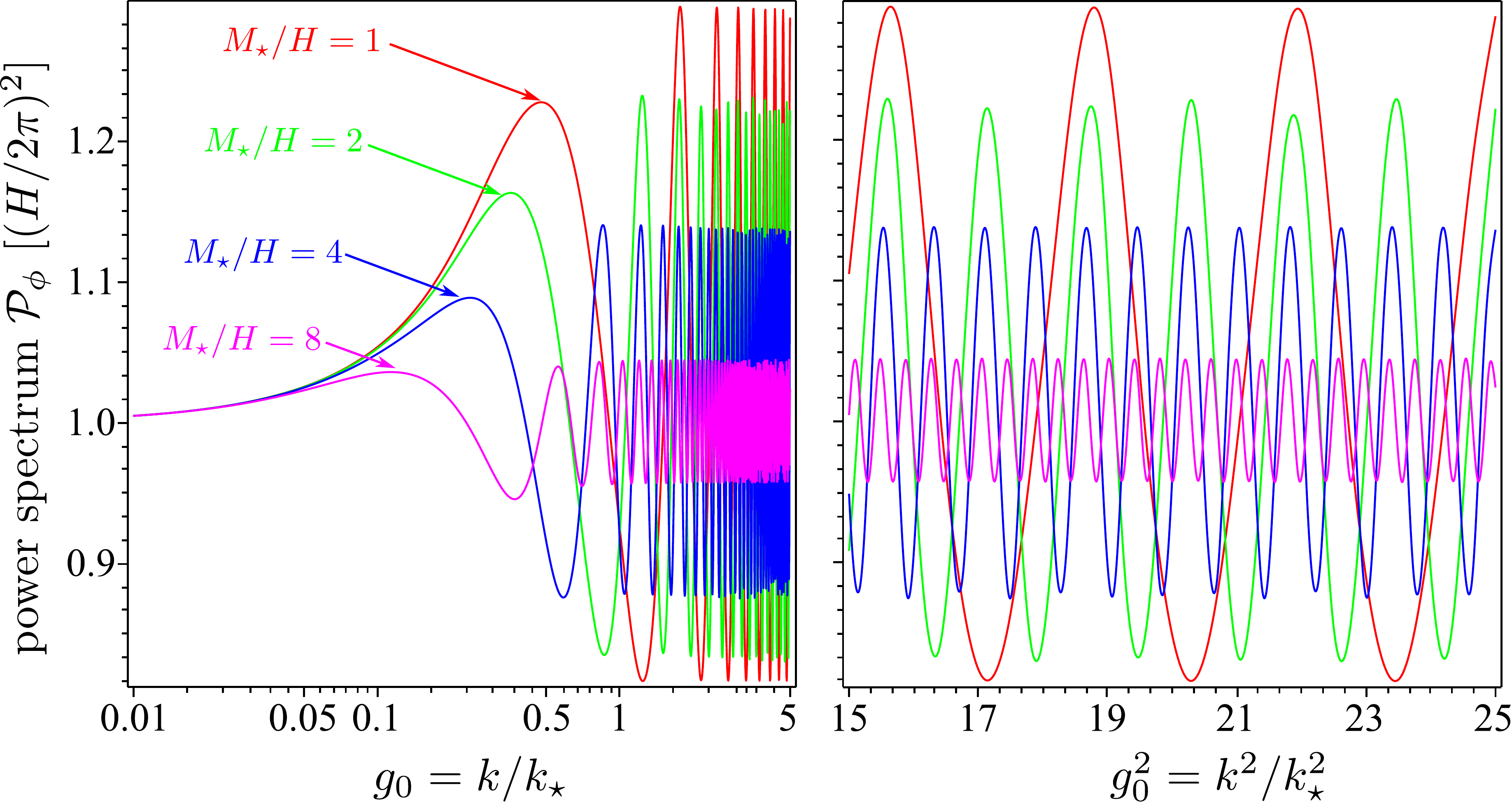}
\end{center}
\caption{Power spectrum of polymer perturbations generated during inflation for various values of $\M/H$.  The righthand panel illustrates that the power spectrum is an oscillatory function of $g_{0}^{2} = k^{2}/k_{\star}^{2}$ for $k \gg k_{\star}$ and that the deviation of the spectrum from the Schr\"odinger result has an amplitude that decreases as $\M/H$ increases.}\label{fig:power spectrum}
\end{figure*}
The numeric simulations described in \S\ref{sec:numeric solutions} can be used to calculate the power spectrum $\mathcal{P}_{\phi}(k)$ using (\ref{eq:polymer power spectrum}) and (\ref{eq:polymer power spectrum 2}).  Results for various choices of $\M/H$ are shown in figure \ref{fig:power spectrum}.  We see that large scale modes with $k \ll k_{\star}$ recover the familiar $\mathcal{P}_{\phi}(k) \approx (H/2\pi)^{2}$ result from Schr\"odinger quantization, while small scale modes with $k \gg k_{\star}$ exhibit an oscillatory power spectrum.  Notice that for $\M/H \gtrsim 4$, the oscillations appear to be sinusoidal.

Using perturbation theory, we have already derived the large scale limit (\ref{eq:large scale polymer power spectrum}) of the polymer power spectrum.  We can also understand the small scale oscillations of the powers spectrum under certain assumptions.  Examining our numerical results for the evolution of a given mode in figure \ref{fig:numeric example}, we see the expansion coefficients closely follow the perturbative prediction up to some transition epoch, and then are roughly constant on large scales.  A crude approximation to this behaviour is to assume that the $g \rightarrow 0$ limit of the expansion coefficients is just given by their perturbative values at the transitional epoch $g = g_\text{tr}$.  More concretely, we can estimate the high $g_{0}$ power spectrum by evaluating the solutions to (\ref{eq:large g EOMs}) at $g = g_\text{tr}$ and then making used of (\ref{eq:polymer power spectrum}) and (\ref{eq:polymer power spectrum 2}).

We can obtain a particularly simple result if we restrict our attention to situations where the argument of the exponential in (\ref{eq:large g EOMs}) is rapidly varying; i.e., $\M \gg H$.  We obtain
\begin{multline}\label{eq:small scale polymer power spectrum}
	\!\!\mathcal{P}_{\phi}(k) = \left( \frac{H}{2\pi} \right)^{2} \left\{ 1 + \frac{5 \sqrt{2}}{768 } \frac{H}{\M g_\text{tr}^{4}} \times \right. \\ \left. \sin \left[ \frac{2\M}{H} \left( \frac{k^{2}}{k^{2}_{\star}} - g_\text{tr}^{2} \right)\right]  + \mathcal{O}\left( \frac{H^{2}}{\M^{2}} \right) \right\}, 
\end{multline}
where this expression is only to be applied when $k \gg k_{\star}$.  The form of this approximate power spectrum is consistent with the simulations results presented in figure \ref{fig:power spectrum}:  The deviation of the polymer power spectrum from the standard result is a sinusoidal function of $k^{2}$ on small scales whose amplitude is inversely proportional to $\M/H$.

Unfortunately, the expression (\ref{eq:small scale polymer power spectrum}) is of little quantitative use without knowing the value of the transition epoch $g_\text{tr}$.  However, we can use the functional form to motivate a fitting formula for our numerical results that is valid on small scales and for $\M \gg H$.  We find that the following expression does a reasonable job of reproducing simulation results for $k/k_{\star}\gtrsim 2$ and $\M/H \gtrsim 50$:
\begin{equation}
	\mathcal{P}_{\phi}(k) \approx \left( \frac{H}{2\pi} \right)^{2} \left\{ 1 +  \frac{H}{4\M} \sin \left[ \frac{2\M}{H} \left( \frac{k^{2}}{k_{\star}^{2}} - 1 \right)\right] \right\}.
\end{equation}

\section{Observational consequences}\label{sec:observations}

In this section, we use the polymer power spectrum derived above to calculate cosmic microwave background (CMB) angular spectra and the present day (linear) matter power spectrum.  Our goal is not a detailed comparison to observations, rather we seek to gain a qualitative understanding of the polymer effects and an indication of whether or not they may be observable.

\subsection{CMB Angular Power Spectrum}
The CMB angular power spectrum provides the highest-quality dataset in modern cosmology, currently best-constrained by the WMAP probe \cite{Komatsu:2010fb}. The angular power spectrum of the temperature auto-correlation is given by
\begin{equation}
C_{\rm{T}l}=\frac{2}{\pi}\int\mathcal{P}(k)\left|\Delta_{\rm{T}l}(k)\right|^2\frac{{d}k}{k}
\end{equation}
where $\Delta_{Tl}(k)$ is the photon temperature transfer function, evaluated at the present epoch. Equivalent forms hold for the $E$-mode polarisation and for the cross-correlation. The transfer functions are typically recovered numerically from a Boltzmann code such as CAMB \cite{Lewis:1999bs} or CLASS \cite{Blas:2011rf}. However, we can first gain insight by focusing on the large scale, small-$l$ region of the temperature auto-correlation, in which
\begin{equation}
\Delta_{\rm{T}l}(k)\approx-\frac{1}{3}j_l[ k(\eta_0-\eta_{\rm dec})] = -\frac{1}{3} j_{l}(kx_{0})
\end{equation}
where $\eta_{\rm dec}$ is the conformal time at decoupling, $\eta_0$ that at the current epoch, and $j_{l}$ is the spherical Bessel function of order $l$. This approximation is valid in Einstein-de Sitter universes and serves as a reasonable approximation for $\Lambda$CDM models with the integrated Sachs-Wolfe effect neglected.  For a scale-invariant (Harrison-Zel'dovich) primordial power spectrum, this transfer function produces the Sachs-Wolfe plateau $l(l+1)C_l=\rm{constant}$, so it could be expected that the polymer quantized primordial power spectrum would produce an approximate Sachs-Wolfe plateau with oscillations imposed upon it. Since the polymer power spectrum tends towards a sinusoid for high wavenumber we expect the impact on smaller scales, and hence higher multipole numbers, to diminish.

This can be demonstrated explicitly for $M_\star/H\gtrsim 5$ and $k_\star\ll k$. Across most of the region of integration, $k>k_\star$, and the approximate form of the power spectrum (\ref{eq:small scale polymer power spectrum}) is a reasonable approximation even for $l=2$, and improving with increasing $l$. Consider the fractional change from a Harrison-Zel'dovich signal,
\begin{multline}
\dfrac{\Delta C_l}{C_l}
=\frac{C_l-C_l^{\rm{HZ}}}{C_l^{\rm HZ}} \\
=\frac{\dfrac{H}{4M_\star}\displaystyle\int\sin\left[\dfrac{2M_\star}{H}\left(\dfrac{k^2}{k_\star^2}-1\right)\right] [j_l(kx_0)]^2\dfrac{{d}k}{k}}{\displaystyle\int [j_l(kx_0)]^2\dfrac{{d}k}{k}}\label{eq:frac change C_l}
\end{multline}
This equation has an exact solution given by a complicated combination of hypergeometric and gamma functions, inducing the expected oscillating Sachs-Wolfe plateau, with the oscillations decaying rapidly. These oscillations are of the order of $2\%$ for $M_\star/H=5$.

The full situation can be solved numerically using the transfer functions produced by the CAMB code and accurate fits for the polymer power spectrum. We take $k_\star=5\times 10^{-4}{\rm Mpc}^{-1}$ and consider the cases $M_\star/H=1$ and $M_\star/H=8$. We employ a flat WMAP7 concordance background with $h=0.704$, $\Omega_bh^2=0.02253$, $\Omega_ch^2=0.1122$, and an amplitude for primordial perturbations of $A_\star=2.48\times 10^{-9}$. However, for simplicity we take a Harrison-Zel'dovich primordial power spectrum with $n_S=1$; the qualitative results and the amplitude of the fractional shifts from the inflationary case are unchanged by the spectral tilt.

\begin{figure}
\includegraphics[width=\columnwidth]{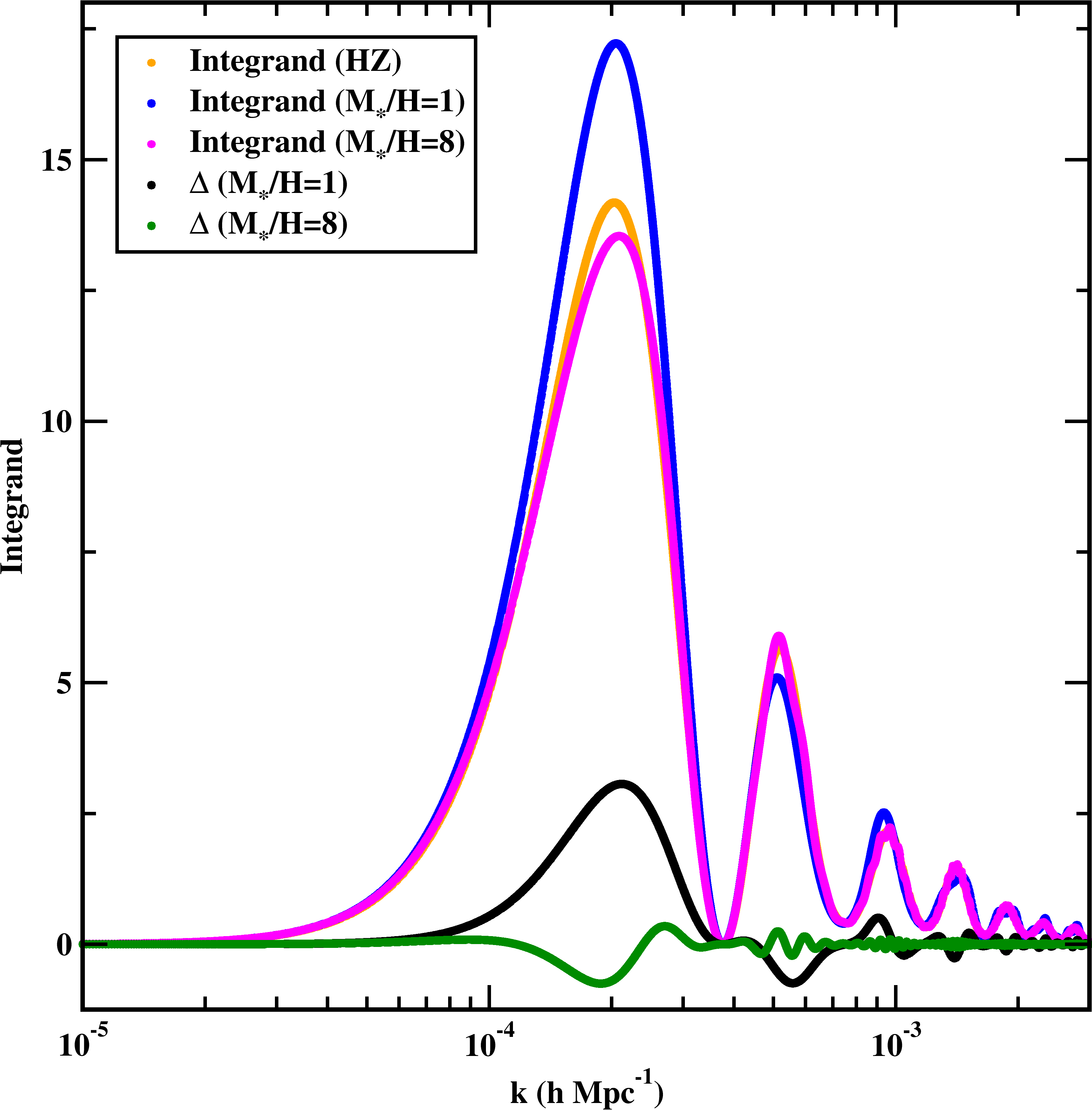}
\caption{The integrands of $C_l$ when $l=2$ for $M_\star/H=1$ and $M_\star/H=8$. Plotted are the integrands for $C_l$ and for $\Delta C_l/C_l$, in comparison with that for the Harrison-Zel'dovich spectrum.}
\label{l2Integrands}
\end{figure}

Figure \ref{l2Integrands} shows the integrands $\mathcal{P}(k)\left|\Delta_{Tl}(k)\right|^2/k$ for the inflationary and polymer models. It is clear that this choice of $k_\star$ provides the best chance of a large impact on the CMB, since the peak in the modulation aligns with the peak in the transfer function at low multipoles.\footnote{It also implies that we might alternatively maximize the impact by either choosing $k_\star$ to align with a peak in the signal at $l=6$, or else to align with a multipole corresponding to the top of the first acoustic peak.}

\begin{figure*}
\includegraphics[width=0.46\textwidth]{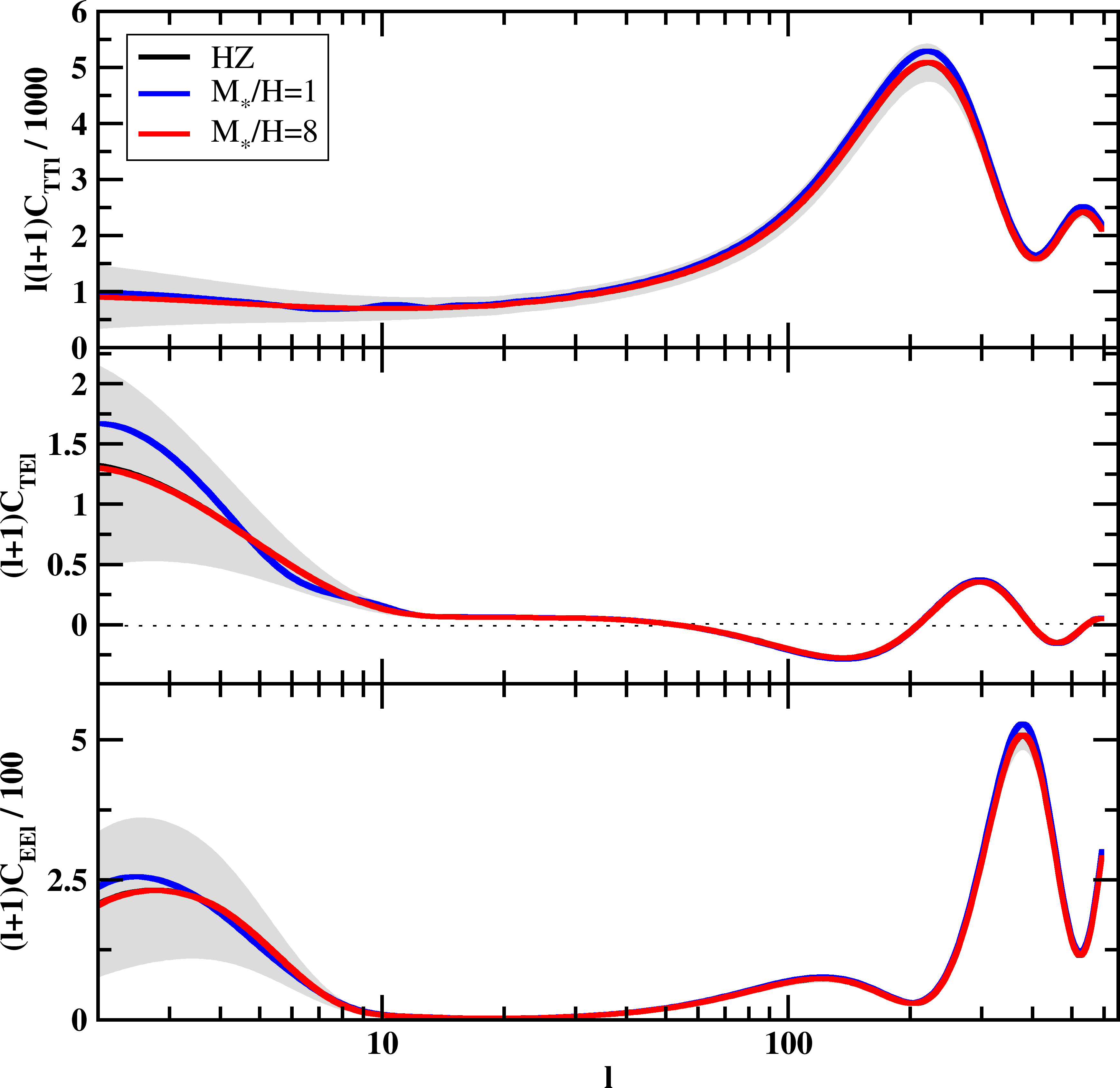}\qquad
\includegraphics[width=0.46\textwidth]{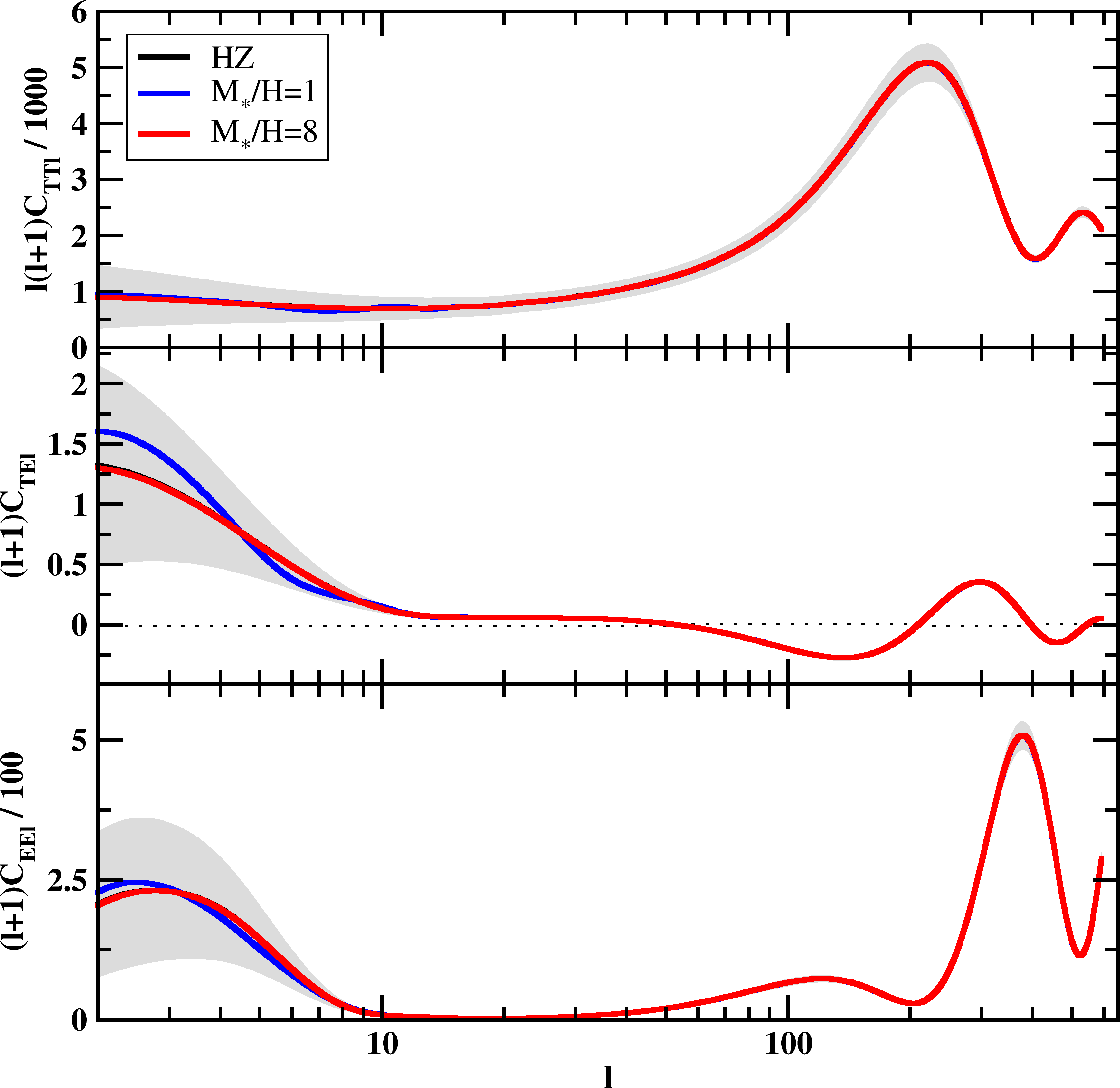}
\caption{The CMB angular power spectra for polymer quantized models with $M_\star/H=1$ (blue) and $M_\star/H=8$ (black), both with $k_\star=5\times 10^{-4}{\rm Mpc}^{-1}$. Left: Equal amplitudes for inflationary and polymer models. Right: Polymer model amplitudes adjusted for high $l$. The grey region shows cosmic variance and the model with $M_\star/H=8$ is indistinguishable by eye from the inflationary case.}
\label{Cl}
\end{figure*}

This setup yields the temperature auto-correlation $C_{TTl}$, the $E$-mode polarization auto-correlation $C_{EEl}$ and the cross-correlation $C_{TEl}$ between these, plotted in Figure \ref{Cl}.

The left panel shows the power spectra assuming that the polymer power spectrum has the same amplitude $A_\star$ as that from inflation. However, since in both models this amplitude is necessarily a free parameter this would be renormalised by the data. The right panel show an amplitude modified such that the deviations from the inflationary signal vanish for higher $l$. Since WMAP is cosmic-variance limited up to the second acoustic peak, and Planck is expected to be cosmic-variance limited up to $l\approx 2000$, this would resemble the best-fit model.

\begin{figure*}
\includegraphics[width=0.46\textwidth]{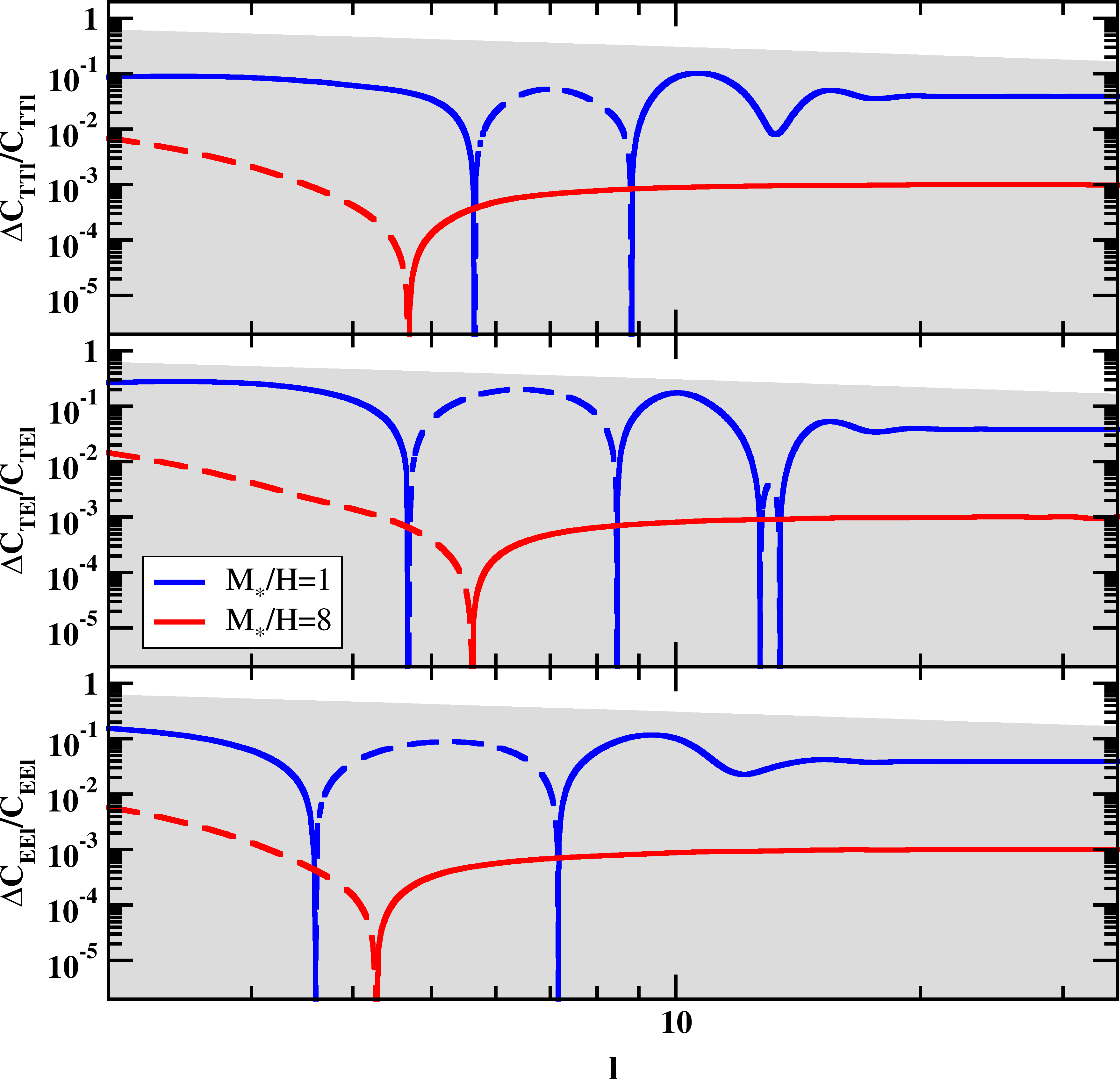}\qquad
\includegraphics[width=0.46\textwidth]{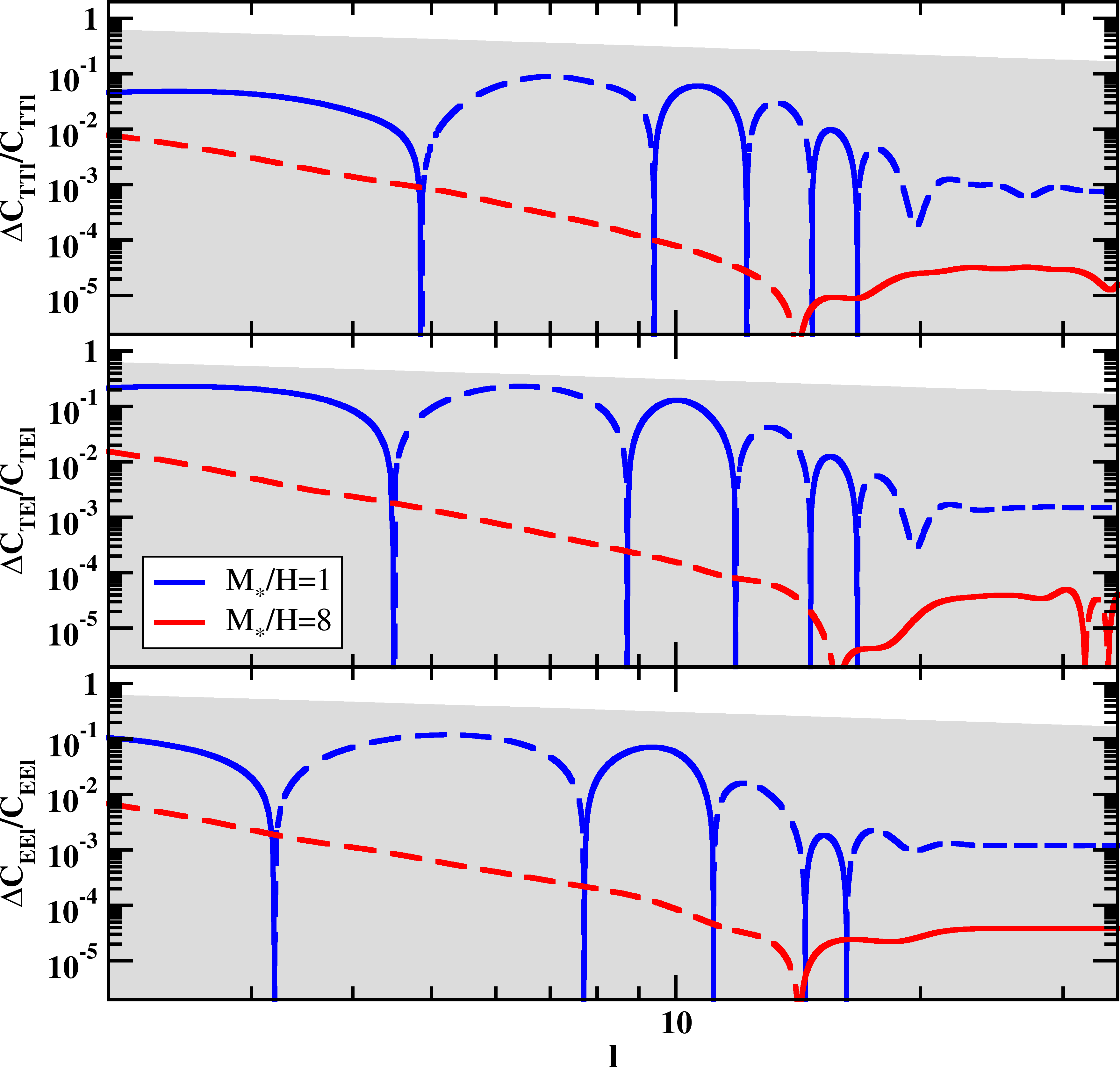}
\caption{The relative shift in the CMB angular power spectra $\Delta C_l/Cl=(C_l-C_l^{\rm HZ})/C_l^{\rm HZ}$ for polymer quantized models with $M_\star/H=1$ (blue) and $M_\star/H=8$ (red), both with $k_\star=5\times 10^{-4}{\rm Mpc}^{-1}$. Left: Equal amplitudes for inflationary and polymer models. Right: Polymer model amplitudes adjusted for high $l$. The grey region shows cosmic variance.  Note that the rapid oscillations in the primordial power spectrum tend to cancel themselves out in the angular power spectra for larger $\M/H$.}
\label{DeltaCl}
\end{figure*}

Figure \ref{DeltaCl} plots the fractional differences from the inflationary model, again both assuming the amplitudes to be the same (left), and adjusting it to match the higher-$l$ signal (right). This plot shows the form of the deviations from the inflationary prediction more clearly.

The CMB angular power spectra for the model with $M_\star/H=8$ are indistinguishable by eye from the inflationary spectra, with a maximum deviation at the quadrupole of $\sim 1\%$, and an asymptotic deviation of $\sim 0.1\%$ even when the amplitude is not renormalised.

The more extreme model with $M_\star/H=1$ causes oscillations in the Sachs-Wolfe regime, with maxima at of the order of $\sim 10\%$ at $l\approx 10$ in each of the correlations, and $l\approx 6$ for the cross-correlation. However, these lie within cosmic variance around the Harrison-Zel'dovich signal and would therefore be extremely difficult to distinguish even for this extreme case. As a result it will be impossible to recover statistically-meaningful constraints from the angular power spectra of the CMB; conversely, the lack of such observed oscillations does not imply that the polymer quantized models are ruled out.

It would be interesting to study the primordial non-Gaussianity induced by the polymer quantized fluctuations; if these are of a characteristic form then it is entirely conceivable that there is a greater observable impact on the CMB bispectrum than on the angular power spectrum. We leave this issue to a future study.


\subsection{Matter Power Spectrum}
The other obvious observable to consider is the matter power spectrum
\be
P_{\rm M}(k,\eta)=\frac{2\pi^2}{k^3}\mathcal{P}(k)\left|\delta(k,\eta)\right|^2 .
\ee

The observable regions of the matter power spectrum are in the regime $k\gg k_\star$, implying that for $M_\star/H\gtrsim 5$ the approximate form of the power spectrum can be used.  The fractional difference between the matter power spectrum in the polymer quantized model and a Harrison-Zel'dovich model is then
\begin{multline}
\frac{\Delta P_\text{M}(k)}{P_\text{M}(k)}=\frac{P_{\rm M}(k)-P_{\rm M,HZ}(k)}{P_{\rm M,HZ}(k)}
\\ \approx\frac{H}{4M_\star}\sin\left[\frac{2M_\star}{H}\left(\frac{k^2}{k_\star^2}-1\right)\right]\sim\frac{H}{4M_\star} .
\end{multline}
We then expect sinusoidal oscillations around the Harrison-Zel'dovich of order $H/4M_\star$, or of the order of 5\% for $M_\star/H\approx 5$, regardless of the value of $k_\star$. The errors on the observed matter power spectrum are currently of the order of $2\%$ \cite{Anderson:2012sa}, implying that a detection may in principle be possible for $M_\star/H\lesssim 10$.  While with current observations it will most likely not be possible to observe the oscillations imprinted on the matter power spectrum, should $M_\star/H$ be low enough it is entirely possible that upcoming missions such as Euclid will observe them.

\begin{figure}
\includegraphics[width=\columnwidth]{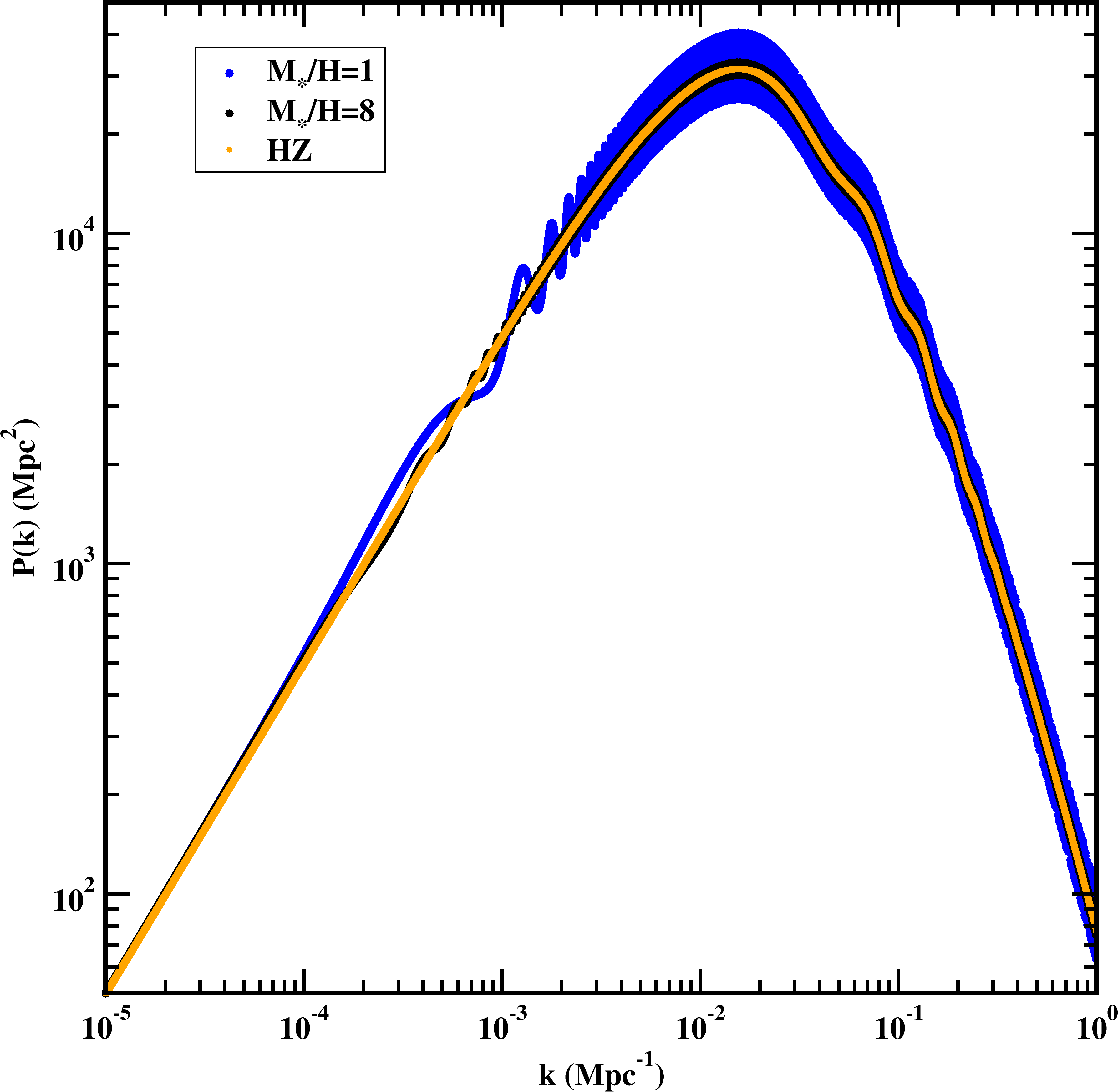}
\caption{The matter power spectrum for a Harrison-Zel'dovich model (orange), and polymer quantized models with $M_\star/H=1$ (blue) and $M_\star/H=8$ (black), both with $k_\star=5\times 10^{-4}{\rm Mpc}^{-1}$.}
\label{MatterPower}
\end{figure}

\begin{figure}
\includegraphics[width=\columnwidth]{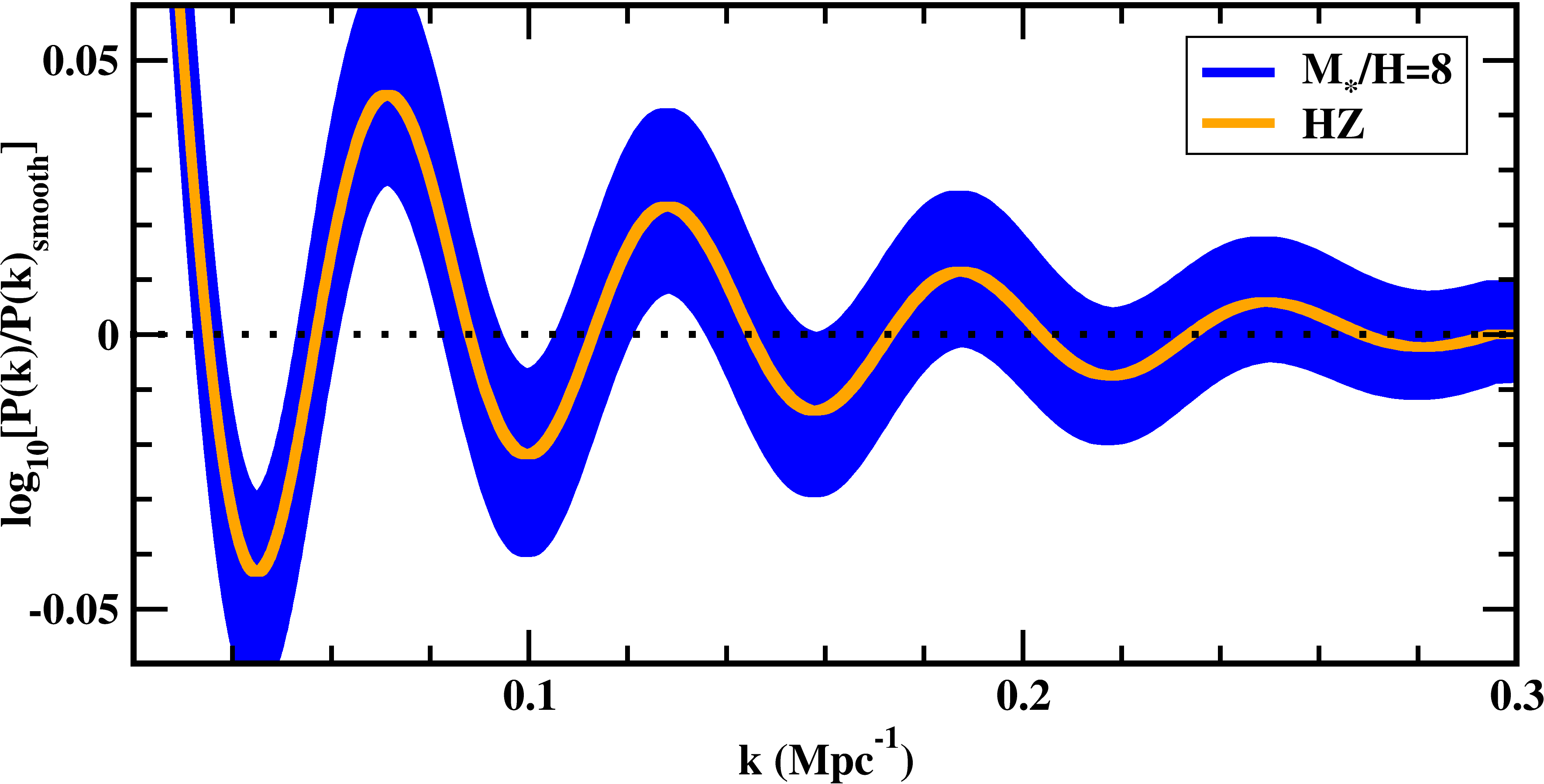}
\caption{Baryon acoustic oscillations for the Harrison-Zel'dovich power spectrum and for the polymer power spectrum with $M_\star/H=8$ and $k_\star=5\times 10^{-4}{\rm Mpc}^{-1}$. This case should be measurable with the BOSS/CMASS survey.}
\label{BAO}
\end{figure}

In Figure \ref{MatterPower} we plot the predicted matter power spectrum for the Harrison-Zel'dovich model, and for the polymer quantized models considered for the CMB. We employ an accurate fit for the primordial power spectrum rather than the approximation employed above. In contrast to the CMB angular power spectrum, where the oscillations induced by the polymer quantisation are integrated out except on very large scales, the matter power spectrum retains these. Figure \ref{BAO} focuses on the baryon acoustic oscillations, plotting the oscillations around a smoothed spectrum. We use a modification of the procedure employed by, for instance, \cite{Percival:2006gs,Anderson:2012sa}, and employ a nonlinear smoothing
\be
\frac{P_{\rm M}(k)}{P_{\rm smooth}(k)}\rightarrow\left[ \frac{P_{\rm M}(k)}{P_{\rm smooth}(k)}-1\right] \exp\left(-\frac{k^2}{\Sigma_\text{NL}^{2}}\right)+1,
\ee
with $\Sigma_{\rm NL}=4.47$ to model the damping of the baryon acoustic oscillations due to nonlinear processes. Plotted are the envelopes of the oscillating spectrum rather than the spectrum itself, and we neglect the case with $M_\star/H=1$; the spread would virtually fill the plot. Such an extreme model is therefore strongly disfavoured by present data. Comparison with Figure 18 in \cite{Anderson:2012sa} suggests that while SDSS-II would be unable to detect the case with $M_\star/H=8$, it is on the edge of detectability with BOSS/CMASS.  The upcoming Euclid probe \cite{Laureijs:2011mu} will be able to constrain the ration $\M/H$ much more tightly.

\section{Discussion}\label{sec:discussion}

In this paper, we have considered the polymer quantization of a massless scalar field in an exactly de Sitter inflationary cosmology.  We have pursued the approach of Fourier transforming the field at the classical level, which reduces the system to a collection of decoupled simple harmonic oscillators, and then numerically solving the resulting time dependent Schr\"odinger equation for the state which closest resembles the Bunch-Davies vacuum at the start of inflation.  We hence calculate the spectrum of primordial perturbations, which recovers the scale-invariant Harrison-Zel'dovich result for $k$ less than a pivot scale $k_{\star}$ corresponding to a mode with physical wavelength $2\pi/\M$ at the beginning of inflation [\emph{cf.}\ (\ref{eq:k star value})].  Here, $\M$ is the polymer energy scale.  For modes with $k \gtrsim k_{\star}$, we find that polymer effects impose oscillations onto the standard result of amplitude $\propto H/\M$.  Numerical plots of the polymer power spectrum are shown in figure \ref{fig:power spectrum} and semi-analytic fitting formulae are given by
\begin{equation}\label{eq:final spectrum}
\frac{\mathcal{P}_{\phi}(k)}{\mathcal{P}_\text{HZ}} =
\begin{cases}
 \displaystyle 1 + \frac{1}{2} \frac{k}{k_{\star}}, & k \ll k_{\star}, \\ \displaystyle
  1 +  \frac{H}{4\M} \sin \left[ \frac{2\M}{H} \left( \frac{k^{2}}{k_{\star}^{2}} - 1 \right)\right], & \begin{array}{l} k \gg k_{\star}, \\ M_{\star} \gg H, \end{array}
 \end{cases}
\end{equation}
where $\mathcal{P}_\text{HZ} = (H/2\pi)^{2}$ is the Harrison-Zel'dovich result.  These are the principal results of this paper.

We have also calculated various CMB power spectra using the polymer power spectrum.  We have seen that the largest effect is for low $l$ multipoles, and even for $H \sim \M$ the magnitude of the effects are within the cosmic variance uncertainty.  It is hence unlikely that the kind of oscillations predicted by this work could be directly observed or constrained in CMB angular spectra.  However, the polymer effects on the present day matter spectra are more pronounced, raising the possibility that future probes of the baryon acoustic oscillations can constrain $H/\M \lesssim 0.1$.

We are prevented by performing a more detailed comparison to observation by the fact that our calculations have been limited to exactly de Sitter cosmologies.  However, we expect our results to go through for slow-roll inflation:  The characteristic inverse timescale associated with polymer effects is given by the logarithmic derivative of the polymer coupling $\tau^{-1}_\text{polymer} = \dot{g}/g = -H$.  On the other hand, the characteristic inverse timsescale associated with changes in the background geometry in slow roll is the logarithmic derivative of the Hubble parameter $\tau^{-1}_\text{slow-roll} = \dot{H}/H$.  Hence we find
\begin{equation}
	\frac{\tau_\text{polymer}}{\tau_\text{slow-roll}} = - \frac{\dot{H}}{H^{2}} = \epsilon \ll 1,
\end{equation}
where $\epsilon$ is one of the standard slow-roll parameters.  Hence, we na\"{\i}vely expect that our polymer power spectrum results to be valid in the slow roll approximation with the $H$ appearing in (\ref{eq:final spectrum}) being interpreted as the Hubble factor at horizon exit for a given mode, but more detailed calculations are warranted.

It is interesting to revisit the model presented in Ref.\ \cite{Hossain:2009ru}, which found that the polymer quantization of a homogeneous scalar could drive quasi-de Sitter inflation with Hubble parameter
\begin{equation}
	H^{2} = \frac{\M^{4}}{12 M^{2}_\text{Pl}}.
\end{equation}
For this model, we then have
\begin{equation}
	\frac{H}{\M} = \frac{1}{\sqrt{6}} \frac{E_\text{inf}}{M_\text{Pl}} = 1.6 \times 10^{-3} \left( \frac{E_\text{inf}}{10^{16}\,\text{GeV}} \right).
\end{equation}
We see that for reasonable choices of the inflationary energy scale, $H/\M$ will be  small for these models.  This implies that polymer effects on the perturbation spectrum will be rather small and likely unobservable.  Furthermore, the slow roll parameters for this model are exponentially small \cite{Hossain:2009ru}, which implies that generated fluctuations will be very close to the Harrison Zel'dovich result.  WMAP data appears to disfavour a purely scale-invariant spectrum \cite{Komatsu:2010fb}, which suggests some tension between polymer-driven inflation and current observations.

Finally, we remark that our technique of Fourier transforming a scalar field at the classical level and then quantizing seems to provide a novel avenue for including high energy/small distance modifications into inflationary cosmology.  It is reasonable to assume that different effects (such as modified uncertainty relations) would lead to different classes of time-dependent potential appearing in the Schr\"odinger equation (\ref{eq:SHO polymer}), and hence different power spectra.  We will report on such models in the future.

\begin{acknowledgements}

SSS would like to thank the Perimeter Institute for Theoretical Physics for hospitality while this work was in progress.  VH and SSS are funded by NSERC of Canada. IAB thanks Bob Nichol, Frode Hansen and Amir Hammami for instructive conversations.

\end{acknowledgements}

\appendix

\section{Polymer quantum mechanics}\label{sec:polymer}

In this appendix, we describe the polymer quantization of a particle moving in the $x$ direction under the influence of a potential $V(x)$.  In \S\ref{sec:PQ}, we will apply the same techniques to quantize the amplitude of a single Fourier mode of a massless scalar during purely de Sitter inflation.

The Hamiltonian of this simple system is
\begin{equation}\label{eq:simple Hamiltonian}
	{\scr H} = \frac{1}{2m} p^{2} + V(x),
\end{equation}
where $p$ is the momentum canonically conjugate to $x$, and we have the Poisson bracket $\{x,p \} =1$.  In polymer quantum mechanics, it is assumed that the quantum state of the system can be expressed as
\begin{equation}\label{eq:arbitrary state}
	|\psi\rangle = \sum^{\infty}_{j=-\infty} c_{j} |x_{j}\rangle.
\end{equation}
Here, $|x_{j}\rangle$ are eigenstates of the position operator
\begin{equation}
	\hat{x} | x_{j} \rangle = x_{j} | x_{j} \rangle, 
\end{equation}
and the $c_{j}$'s are expansion coefficients.  Note that we will assume that all the time dependence of the system is carried in the state vector $|\psi\rangle = |\psi\rangle (t)$ (i.e., we take the Schr\"odinger picture), which means the expansion coefficients will in general depend on time $c_{j} = c_{j}(t)$.  The spectrum of the position operator $\left\{ x_{j} \right\}$ consists of a countable selection of points from the real line $\mathbb{R}$ which is analogous to the graphs covering 3-manifolds in LQG.  A critical difference between polymer and Schr\"odinger quantizations is that in the former, the position eigenstates are normalizable:
\begin{equation}
	\langle x_{i} | x_{j} \rangle = \delta_{i,j}.
\end{equation}
A translation operator can be defined by it's action on position eigenstates:
\begin{equation}
	\hat{U}_{\lambda} |x_{j}\rangle = |x_{j}-\lambda\rangle;
\end{equation} 
that is, $\hat{U}_{\lambda}$ converts a position eigenstate with eigenvalue $x_{j}$ into an eigenstate with eigenvalue $x_{j}-\lambda$.  One can easily confirm the commutator identity
\begin{equation}
	[\hat{x},\hat{U}_{\lambda}] = -\lambda \hat{U}_{\lambda},
\end{equation}
by acting the lefthand side on an arbitrary state (\ref{eq:arbitrary state}).

In Schr\"odinger quantum mechanics, the relationship between the translation and momentum operators is $\hat{U}_{\lambda} = e^{i\lambda\hat{p}}$; i.e., $\hat{p}$ is the generator of infinitesimal translations.  Another way to state this is:
\begin{equation}\label{eq:Schrodinger momentum definition}
	\hat{p} \equiv -i  \frac{\di \hat{U}_{\lambda}}{\di \lambda} \bigg|_{\lambda=0}.
\end{equation}
Hence in order to define a momentum operator, $\hat{U}_{\lambda}$ must be differentiable at $\lambda = 0$.  It is fairly easy to see that this is not the case: for example, consider the expectation value of $\hat{U}_{\lambda}$ in a position eigenstate:
\begin{equation}
	\langle x_{i} | \hat{U}_{\lambda} | x_{i} \rangle = \begin{cases} 1, & \lambda = 0, \\ 0, & \lambda \ne 0. \end{cases}
\end{equation}
In is clear that the matrix representation of $\hat{U}_\lambda$ in the $\{|x_{i}\rangle\}$ basis (and hence $\hat{U}_{\lambda}$ itself) fails to be differentiable at $\lambda = 0$.  Therefore, we cannot define a momentum operator in the polymer setup.  

However, we can define an effective momentum operator by creating a finite difference stencil of the definition (\ref{eq:Schrodinger momentum definition}):
\begin{equation}\label{eq:p star}
	\hat{p}_{\lambda_{\star}} \equiv -i \left(\frac{\hat{U}_{\lambda_{\star}}-\hat{U}_{-\lambda_{\star}}}{2\lambda_{\star}} \right) = \frac{\hat{U}_{\lambda_{\star}}-\hat{U}^{\dag}_{\lambda_{\star}}}{2i\lambda_{\star}}.
\end{equation}
(This is not the only way to define $\hat{p}_{\lambda_{\star}}$, see footnote \ref{foot:ambiguity} for more details.)  The fixed width of our finite difference stencil $\lambda_{\star}$ plays a crucial role in this quantization scheme: As mentioned above, it defines and energy scale $\M$ above which Schr\"odinger and polymer predictions will diverge.\footnote{The precise relationship between $\lambda_{\star}$ and $\M$ depends on the dimensions of $x$: If $x$ is a length, then $\M = 1/\lambda_{\star}.$}  Practical calculations in polymer quantum mechanics involve mapping $\hat{p} \mapsto \hat{p}_{\lambda_{\star}}$ in the operator version of the classical Hamiltonian (\ref{eq:simple Hamiltonian}):
\begin{multline}\label{eq:simple Hamiltonian}
	\hat{\!\! \scr H} = \frac{1}{2m} \hat{p}_{\lambda_{\star}}^{2} + V(\hat{x})  = \\ \frac{1}{8m\lambda_{\star}^{2}} (2- \hat{U}_{2\lambda_{\star}} - \hat{U}^{\dag}_{2\lambda_{\star}} )+ V(\hat{x}).
\end{multline}
Armed with this operator representation of the Hamiltonian, we can attempt to solve the Schr\"odinger equation
\begin{equation}\label{eq:simple Schrodinger}
	i \di_{t} |\psi \rangle = \, \hat{\!\! \scr H}  |\psi \rangle,
\end{equation}
and hence determine the time evolution of observable feature of our system.

Now, the graph $\{ x_{j} \}$ involved in our decomposition of a generic state vector $|\psi\rangle$ is meant to be an arbitrary selections of point on the real line, not necessarily a regularly spaced lattice.  However, we find that the Hamiltonian tends to ``super-select'' regular lattices in the following sense:  We say that a state vector is in the subspace $\mathcal{H}_{\text{poly}}^{x_{0}}$ of the entire polymer Hilbert space $\mathcal{H}_{\text{poly}}$ if it can be expressed as
 \begin{equation}\label{eq:superselected state}
	|\psi\rangle = \sum^{\infty}_{j=-\infty} c_{j} |x_0 + 2\lambda_{\star} j \rangle \quad \Leftrightarrow \quad |\psi \rangle \in \mathcal{H}_{\text{poly}}^{x_{0}},
\end{equation}
where $\lambda_{\star}$ is the fixed scale we used in the definition of $\hat{p}_{\lambda_{\star}}$ (\ref{eq:p star}).  It then follows that if $|\psi \rangle$ is in $\mathcal{H}_{\text{poly}}^{x_{0}}$ then the Hamiltonian (\ref{eq:simple Hamiltonian}) acting on $|\psi \rangle$ is also in $\mathcal{H}_{\text{poly}}^{x_{0}}$; i.e.,
\begin{equation}
       |\psi \rangle \in \mathcal{H}_{\text{poly}}^{x_{0}} \quad \Rightarrow \quad \hat{\!\!\scr H} |\psi \rangle \in \mathcal{H}_{\text{poly}}^{x_{0}}.
\end{equation}
Hence, the Schr\"odinger equation (\ref{eq:simple Schrodinger}) and the associated eigenvalue problem can be solved in each of the super-selected sectors separately.  In more practical terms, this means that it sufficient to assume state vectors of the form (\ref{eq:superselected state}) in (\ref{eq:simple Schrodinger}), which translates the problem into an infinite dimension matrix ODE for the expansion coefficients $\mathbf{c} = [c_{j}(t)]$.

The solution of such a problem can be unwieldy, so we introduce an alternative representation.  Let us define a new basis via
\begin{equation}\label{eq:momentum basis definition}
	|p\rangle = \sum_{j=-\infty}^{\infty} e^{-ipx_{j}} |x_{j}\rangle, \quad \langle p | x_{j} \rangle = e^{ipx_{j}}.
\end{equation}
An arbitrary state in this basis is expressed as $\psi(p) = \langle p | \psi \rangle$.  It follows that
\begin{equation}\label{eq:momentum space wavefunction}
	|\psi\rangle = \sum^{\infty}_{j=-\infty} c_{j} |x_{j}\rangle \quad \Rightarrow \quad \psi(t,p) = \sum_{j=-\infty}^{\infty} c_{j} e^{ipx_{j}}.
\end{equation}
In Schr\"odinger quantum mechanics, $|p\rangle$ would be an unnormalizable momentum eigenstate with the above sums replaced by integrals; however, the lack of a momentum operator in polymer quantization prevents us from making a similar interpretation here.  Nevertheless, we call $\psi(t,p)$ the momentum-space wavefunction.  It is easy to deduce the action of fundamental operators in this representation:
\begin{equation}
	\langle p | \hat{x} | \psi \rangle = i \frac{\di}{\di p} \psi(t,p), \quad \langle p | \hat{U}_{\lambda} | \psi \rangle = e^{ip\lambda} \psi(t,p).
\end{equation}
In this basis, the Schr\"odinger equation \ref{eq:simple Schrodinger} becomes a differential equation:
\begin{equation}
	i \frac{\di}{\di t} \psi(t,p) = \left[ \frac{ \sin^{2} (\lambda_{\star} p)}{2m\lambda_{\star}^{2}} + V\left(  -i \frac{\di}{\di p} \right) \right] \psi(t,p).
\end{equation}
Whether or not this is easier to solve than the difference equation implicitly defined by (\ref{eq:simple Schrodinger}) depends on the form of the potential.

Let us now restrict ourselves to a particular super-selected subspace $\mathcal{H}_{\text{poly}}^{x_{0}}$.  An identity operator on such a subspace is
\begin{equation}
	\hat{\mathbf 1} = \int\limits_{-\pi/2\lambda_{\star}}^{\pi/2\lambda_{\star}} dp \, |p\rangle \langle p|.
\end{equation}
It is easy to confirm that if $|\psi \rangle \in \mathcal{H}_{\text{poly}}^{x_{0}}$ and $|\phi \rangle \in \mathcal{H}_{\text{poly}}^{x_{0}}$, then
\begin{equation}
	\langle \phi | \psi \rangle = \langle \phi | \hat{\mathbf 1} | \psi \rangle = \int\limits_{-\pi/2\lambda_{\star}}^{\pi/2\lambda_{\star}} \phi^{*}(t,p) \psi(t,p) \, dp,
\end{equation}
which serves to define the inner product on $\mathcal{H}_{\text{poly}}^{x_{0}}$ in the momentum representation.

Finally, we note that if we restrict ourselves to states in $\mathcal{H}_{\text{poly}}^{x_{0}}$, we can take $p \in [-\pi/2\lambda_{\star},\pi/2\lambda_{\star}]$.  From the definition (\ref{eq:momentum space wavefunction}) and $x_{j}= x_{0}+2\lambda_{\star}j$, the momentum space wavefunction must satisfy the following boundary conditions:
\begin{equation}\label{eq:polymer BC} 
	\psi(t,-\pi/2\lambda_{\star}) = e^{i\pi x_{0}/\lambda_{\star}}\psi(t,\pi/2\lambda_{\star}). 
\end{equation}
A specific choice of lattice offset will result in a specific boundary condition:  For example, if we choose $x_{0} = 0$, the wavefunction will be periodic with period $\pi/\lambda_{\star}$, while the choice $x_{0} = \lambda_{\star}$ results in an anti-periodic wavefunction.  This approach to selecting boundary conditions will have physical ramifications; in the case of the polymer simple harmonic oscillator, the energy eigenvalues will depend on $x_{0}$ \cite{Ashtekar:2002sn,Hossain:2010eb}.  In this paper, we take a different approach by demanding $\psi(t,-\pi/2\lambda_{\star}) = \psi(t,\pi/2\lambda_{\star}) = 0$.  In, \S\ref{sec:PQ formal} we argue that such a choice is required to ensure unitary evolution with a time-dependent $\lambda$.  Here, we see a separate justification:  This is the only choice of boundary condition that will ensure physical observables do not depend on $x_{0}$; i.e., the choice of regular lattice in which we perform calculations.

\section{Scaling formulae}\label{sec:scaling}

In an FRW universe, all physical or observable quantities must be invariant under dilations of the spatial coordinates, or equivalently, re-scalings of the scale factor $a$.  It is useful to make note of how the various quantities defined above transform under a uniform scaling of the spatial coordinates by a constant factor $\ell$:
\begin{align}\label{eq:scalings}
	\nonumber a & \rightarrow \ell^{-1} a , & \x & \rightarrow \ell \x, & V_{0} & \rightarrow \ell^{3} V_{0}, \\
	\nonumber \phi & \rightarrow \phi, & \pi & \rightarrow \ell^{-3} \pi, & \k & \rightarrow \ell^{-1} \k, \\
	\phi_{\k} & \rightarrow \ell^{3/2} \phi_{\k}, & \pi_{\k} & \rightarrow \ell^{-3/2} \pi_{\k}.
\end{align}
We use these relations in \S\ref{sec:PQ} to define field amplitude translation and effective polymer momentum operators that transform correctly under dilations $\mathbf{x} \rightarrow \ell \mathbf{x}$.

\section{Derivation of $k_{\star}$}\label{sec:k star derivation}

Most of our calculations and results are defined in terms of a quantity $k_{\star}$ which is the present wavenumber of a Fourier mode whose physical wavenumber was equal to $\M$ at the start of inflation.  In this appendix, we derive an explicit formula for $k_{\star}$ in terms of familiar inflationary parameters.  Define $a_\text{now}$ as the current value of the scale factor, $a_\text{end}$ as the value of the scale factor at the end of inflation, and $a_\text{start}$ to be the scale factor at the start of inflation.  Then,
\begin{equation}
	k_{\star} = \frac{a_\text{end}}{a_\text{now}} \frac{a_\text{start}}{a_\text{end}} \M.
\end{equation}
The number of $e$-folds of inflation is
\begin{equation}
	N = \ln \frac{a_\text{end}}{a_\text{start}}.
\end{equation}
The radiation-like component of the universe's total density evolves adiabatically after inflation.  Equating the entropy in the radiation fluid now with the entropy just after the end of inflation yields
\begin{equation}
	\mathcal{G}_\text{end} a_\text{end}^{3} T_\text{end}^{3} = \mathcal{G}_\text{now} a_\text{now}^{3} T_\text{now}^{3}, 
\end{equation}
where $\mathcal{G}$ is the effective number of relativistic species in the universe at a given epoch, and $T$ is the temperature.  We can use the Stepan-Boltzman law to relate the temperature to the total density of radiation at the end of inflation:
\begin{equation}
	\rho_\text{end} = \frac{1}{30} \mathcal{G}_\text{end} \pi^{2} k_\text{b}^{4} T_\text{end}^{4},
\end{equation} 
Where $k_\text{b}$ is Boltzmann's constant.  Just after reheating, the universe is radiation dominated so $\rho_\text{end}$ is actually the density appearing in the Friedman equation:
\begin{equation}
	H^{2}_\text{end} = H^{2} = \frac{\rho_\text{end}}{3M_\Pl^{2}} = \frac{E_\text{inf}^{4}}{3M_\Pl^{2}},
\end{equation}
where $H$ is the Hubble parameter during inflation and $E_\text{inf} = \rho^{1/4}_\text{end}$ is the inflationary energy scale.  Assembling these results, we obtain
\begin{equation}
	k_\star = \frac{\pi^{1/2}k_\text{b}T_\text{now} \mathcal{G}_\text{now}^{1/3}}{3 \cdot 30^{1/4}} \mathcal{G}_\text{end}^{-1/12} e^{-N} \frac{\M}{H} \frac{E_\text{inf}}{M_\Pl}.
\end{equation}
Using $T_\text{now} = 3.94\,\text{K}$ and $\mathcal{G}_\text{now} = 3.04$, we obtain (\ref{eq:k star value}).

\bibliographystyle{arxiv_physrev}
\bibliography{poly_spec}

\end{document}